%% file: paper.tex
\documentclass[a4paper,fleqn,useAMS,usenatbib]{mnras}
\usepackage{graphicx}
\usepackage{pdflscape}  

\newcommand{\teff}{$T_{\rm eff}$}

\title[{\it Spitzer} observations of NGC 6822]{\textit{Spitzer} observations of extragalactic H~{\sc ii} regions -- III. NGC~6822 and the hot star,  H~{\Large \textbf{II}} region connection}

\author[R. H. Rubin et al.]
{Robert H. Rubin$^{1,2}$, 
Janet P. Simpson$^{3}$\thanks{E-mail: janet.p.simpson@gmail.com},
Sean W. J. Colgan$^{1}$,
Reginald J. Dufour$^{4}$,
\newauthor
Justin Kader$^{1,5}$, 
Ian A. McNabb$^{1,6}$,
Adalbert W. A. Pauldrach$^{7}$,
Johann A. Weber$^{7}$ \\ 
$^{1}$NASA/Ames Research Center, Moffett Field, CA 94035-1000, USA\\
$^{2}$Orion Enterprises; deceased 3 March 2013 \\
$^{3}$SETI Institute, 189 Bernardo Ave, Mountain View, CA 94043, USA \\
$^{4}$Physics \& Astronomy Department, Rice University, MS 108, Houston, TX 77005-1892, USA \\
$^{5}$Department of Physics \& Astronomy, San Francisco State University, 1600 Holloway Ave, San Francisco, CA 94132, USA  \\
$^{6}$Kavli Institute for Astronomy and Astrophysics, Peking University, Beijing 100871, China \\
$^{7}$Universitaets-Sternwarte Muenchen, Scheinerstrasse 1, D-81679 Muenchen, Germany \\
}

\begin{document}

\date{\today}

\maketitle

\label{firstpage}

\begin{abstract}

Using the short-high module of the Infrared Spectrograph on the {\it Spitzer Space Telescope}, 
we have measured the [S~{\sc iv}] 10.51, [Ne~{\sc ii}] 12.81, [Ne~{\sc iii}] 15.56, and [S~{\sc iii}] 18.71-$\mu$m emission lines in nine H~{\sc ii} regions in the dwarf irregular galaxy NGC~6822. 
These lines arise from the dominant ionization states of the elements neon (Ne$^{++}$, Ne$^+$) and sulphur (S$^{3+}$, S$^{++}$), thereby allowing an analysis of the neon to sulphur abundance ratio as well as the ionic abundance ratios Ne$^+$/Ne$^{++}$ and S$^{3+}$/S$^{++}$. 
By extending our studies of H~{\sc ii} regions in M83 and M33 to the lower metallicity NGC~6822, we increase the reliability of the estimated Ne/S ratio. 
We find that the Ne/S ratio appears to be fairly universal, with not much variation about the ratio found for NGC~6822: the median (average) Ne/S ratio equals 11.6 (12.2$\pm$0.8). 
This value is in contrast to Asplund et al.'s currently best estimated value for the Sun: Ne/S = 6.5. 
In addition, we continue to test the predicted ionizing spectral energy distributions (SEDs) 
from various stellar atmosphere models by comparing model nebulae computed with these SEDs as inputs to our observational data, changing just the stellar atmosphere model abundances. 
Here we employ a new grid of SEDs computed with different metallicities: Solar, 0.4 Solar, and 0.1 Solar. 
As expected, these changes to the SED show similar trends to those seen upon changing just the nebular gas metallicities in our plasma simulations: lower metallicity results in higher ionization. 
This trend agrees with the observations.

\end{abstract}


\section{Introduction}

In this paper we continue our study of extragalactic H~{\sc ii} regions using the {\it Spitzer Space Telescope} (Werner et al. 2004). 
Previously we had taken measurements of emission lines in 24 H~{\sc ii} regions 
in the substantially face-on spiral galaxy M83 (Rubin et al. 2007, hereafter R07) 
that we observed with {\it Spitzer} and of 25 H~{\sc ii} regions in the Local Group 
spiral galaxy M33 (Rubin et al. 2008, hereafter R08), also observed with {\it Spitzer}. 
Here we combine our measurements
and derived quantities from those studies with {\it Spitzer} data from 9 H~{\sc ii} regions
in the Local Group dwarf irregular galaxy NGC 6822 in order to more fully examine
trends that resulted from our previous analysis. 
This is a stand-alone paper, as a consequence some material from R07 and R08 will be repeated here.

An effective means of studying the chemical evolution of the universe is studying emission line objects. 
Different types of emission nebulae are characteristic of different evolutionary stages, 
and are comprised of various mixes of elemental abundances. 
Here, we address two of the abundant elements with observations of extragalactic gaseous nebulae in 
order to probe for trends in galactic chemical evolution (GCE), 
as well as the current state of the interstellar medium (ISM). 
The study of gaseous nebulae refines our understanding of the physical processes of stellar, galactic and 
primordial nucleosynthesis. 
Using {\it Spitzer}, we have a means of reliably  studying the elements neon and sulphur. 
In H~{\sc ii} regions, the dominant ionization states of these elements, 
Ne (Ne$^+$, Ne$^{++}$) and S (S$^{++}$, S$^{3+}$) are all present and 
their four corresponding emission lines can be observed simultaneously. 
Using the Infrared Spectrograph (IRS, Houck et al. 2004) aboard {\it Spitzer}, 
we can observe the four lines 
[S~{\sc iv}] 10.51,
[Ne~{\sc ii}] 12.81,
[Ne~{\sc iii}] 15.56,
and [S~{\sc iii}] 18.71~$\mu$m cospatially.
The ability to observe these lines simultaneously and cospatially, 
combined with the extreme sensitivity of the IRS, 
makes one of {\it Spitzer's} specialties the study of extragalactic H~{\sc ii} regions. 

With this in mind, we have studied prominent H~{\sc ii} regions 
in galaxies with varying metallicity, ionization, and morphology. 
In taking such a survey of the dominant ionization states of neon and sulphur, 
we can infer a Ne/S ratio. 
We also have to consider a contribution from S$^+$, as measurements of any emission lines 
corresponding to this ionization state were not made. 
Some non-ionic S is also expected to be present, but tied up in dust grains and molecular gas. 
As such, we surmised that our values for Ne/S for M83 were in fact upper limits. 
Due to its relatively low ionization, 
M83 would have larger contributions from S$^+$, and less from S$^{++}$ and S$^{3+}$. 
We suspected (R07) this is to blame for the large Ne/S values inferred from the  H~{\sc ii} regions in M83. 
We concluded (R08) that the lower Ne/S values derived for the M33
H~{\sc ii} regions are more reliable estimates because M33 has lower metallicity and higher ionization than M83. 
Our choice of further observations of H~{\sc ii} regions  in NGC 6822 
was motivated by its low metallicity and high ionization.  

In both gaseous nebulae and stars, the presence of radial (metal/H) abundance gradients in the 
plane of the Milky Way 
has been repeatedly observed (e.g. Henry \& Worthey 1999; Rolleston et al. 2000). 
These radial abundance gradients seem to be common to all
spiral galaxies, though the degree varies depending on a given spiral's morphology and luminosity class. 
We observed a trend of higher (Ne/H) and (S/H) with lower deprojected galactocentric radii ($R_{\rm G}$) in M33. 
These gradients are usually attributed to the radial dependence of star formation history and ISM mixing processes (e.g. Shields 2002)
as affected by galaxy outflows (winds) and gas accretion  
(see the review of S\'anchez Almeida et al. 2014). 
Thus, the observed gradients are another tool for understanding galactic evolution 
(e.g. Hou, Prantzos \& Boissier 2000; Chiappini, Matteucci \& Romano 2001; Chiappini, Romano \& Matteucci 2003; Kubryk, Prantzos \& Athanassoula 2015). 
The concept is that star formation and chemical enrichment 
begin in the nuclear bulges at the center of the galaxies and thereafter radially progress outward, 
where much gas remains.

When comparing data to various nebular plasma simulations, 
it has become evident that the results from the
photoionization modeling codes are highly sensitive to the ionizing spectral energy distribution (SED) 
that is input (e.g. R08, R07, Simpson et al. 2004, and references therein). 
These input SEDs come from stellar 
atmosphere models, and in R07 we developed new observational tests of and constraints on the ionizing SEDs that are predicted 
by these stellar atmosphere models. 
The fits mentioned above compare our {\it Spitzer} observations of H~{\sc ii} regions in M83 and M33 with 
predictions from our photoionization simulations that vary {\it only} the ionizing SED.
M83 provided data for high metallicity (at least twice solar, e.g. Dufour et al. 1980; Bresolin \& Kennicutt 2002) and 
lower ionization H~{\sc ii} regions. 
M33 along with N66 (SMC) and 30 Dor (LMC) have provided data for relatively
lower metallicity and higher ionization H~{\sc ii} regions. 
Previously our data for the most part were best fit with nebular models  having SEDs predicted by supergiant stellar 
atmosphere models calculated with the {\sc wm-basic} code (Pauldrach, Hoffmann \& Lennon 2001; Sternberg, Hoffman \& Pauldrach 2003).
The current paper now allows for a more detailed investigation, now including NGC 6822.

In Section 2 we describe the NGC 6822 {\it Spitzer}/IRS observations. 
In Section 3 we discuss the computation of ionic abundance ratios, 
particularly as they apply to NGC 6822.
Section 4 describes how our {\it Spitzer} data are used to 
constrain and test the ionizing SEDs predicted by stellar atmosphere models. 
In our Section 5 we present the Ne/S abundance ratios for our new NGC 6822 \mbox{H\,{\sc ii}} region observations 
and in addition, 
discuss the abundance ratios as functions of \mbox{H\,{\sc ii}} region excitation and metallicity.
Last, in Section 6, we provide a summary and conclusions.
Because there have been significant recent advances in the computation 
of effective collision strengths, which are used in the determination of abundances,
we recalculate the ionic abundance and total abundance ratios 
from the data in previous papers. 
These results on M42 (Rubin et al. 2011, hereafter R11), R07, R08, and other galaxy data 
from Wu et al. (2008, hereafter W08) and 
Lebouteiller et al. (2008, hereafter L08) and discussed in R08 
are found in the Appendix.

\section{\textit{Spitzer Space Telescope} Observations}


\input table1.tex


\begin{figure}
\resizebox{8.5cm}{!}{\includegraphics{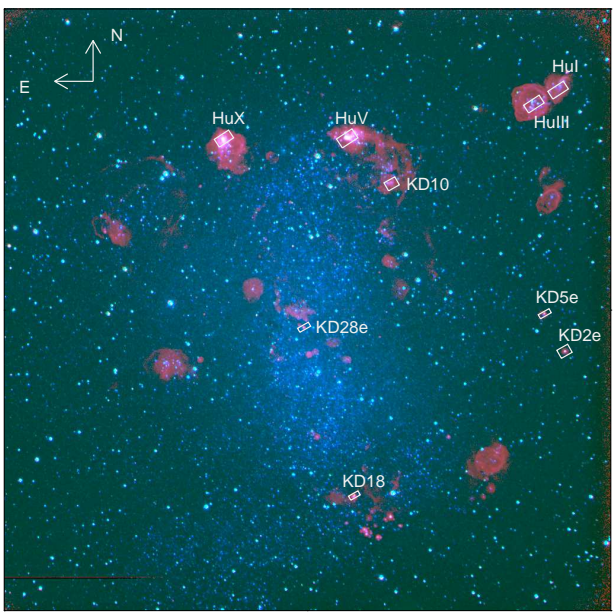}} 
\caption[]{Three-colour image of NGC 6822 showing the locations of the observed H~{\sc ii} regions and the apertures used for extracting the spectra.
The images were taken for the Spitzer Infrared Nearby Galaxies Survey (SINGS, Kennicutt et al. 2003) and were obtained from the NASA/IPAC Extragalactic Database (NED;
Helou et al. 1991).
The colours are as follows: blue -- B band, green -- V band, and red -- H$\alpha$.
North is up.
}
\end{figure}


\begin{figure*}
    \begin{tabular}{cc}
\resizebox{7cm}{!}{\includegraphics{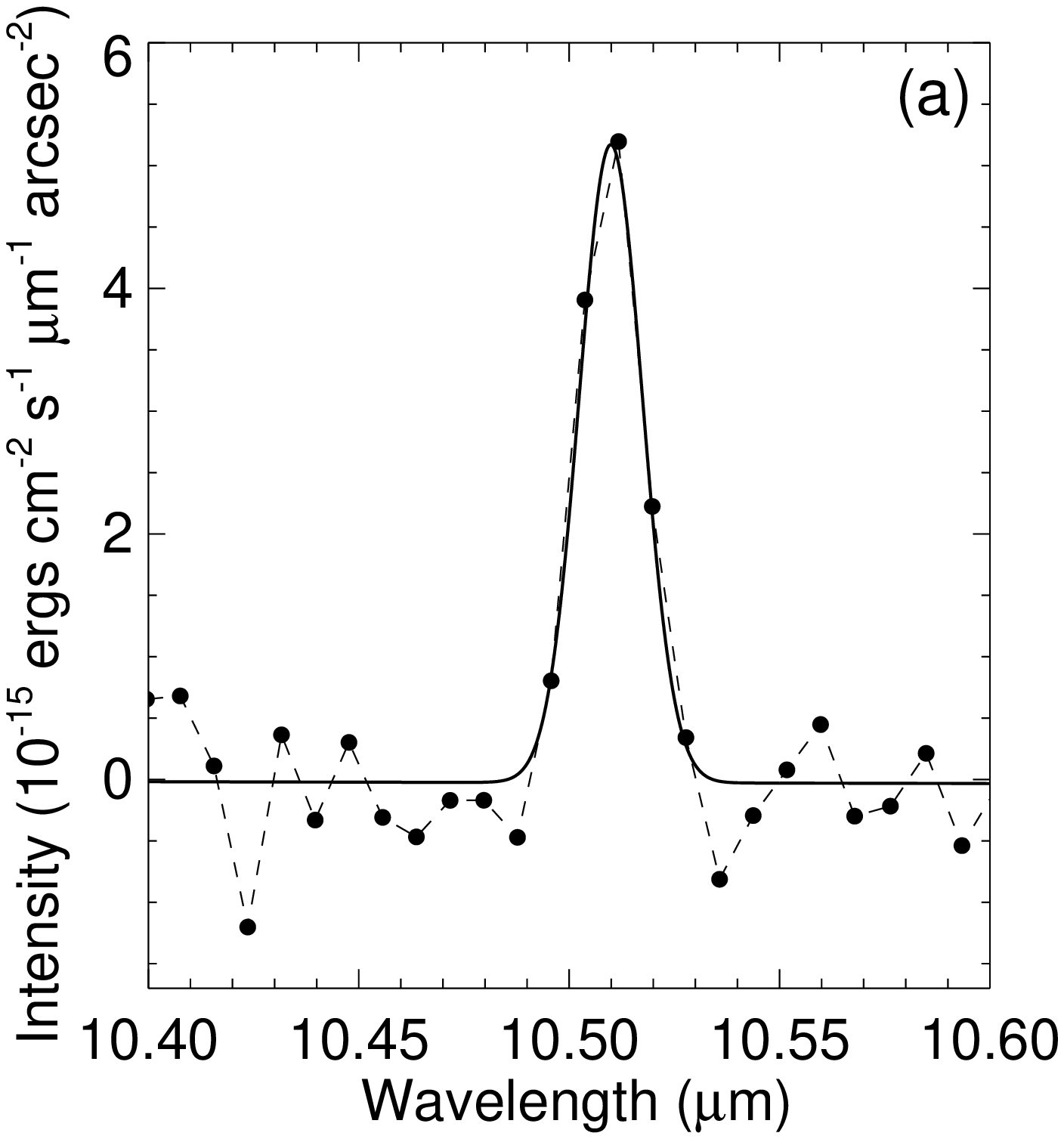}} &
\resizebox{7cm}{!}{\includegraphics{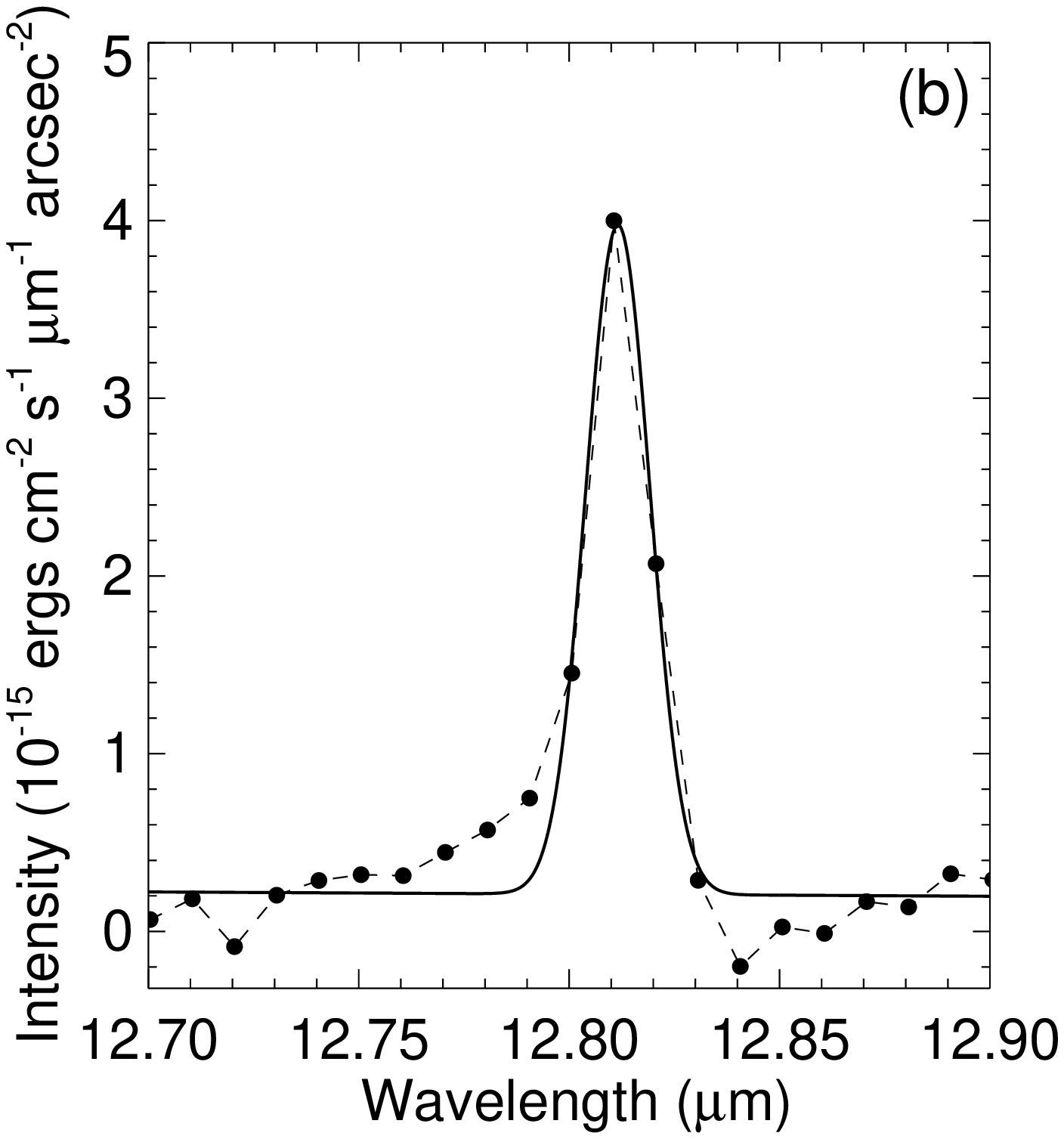}} \\
	\end{tabular}
	\begin{tabular}{cc}
\resizebox{7cm}{!}{\includegraphics{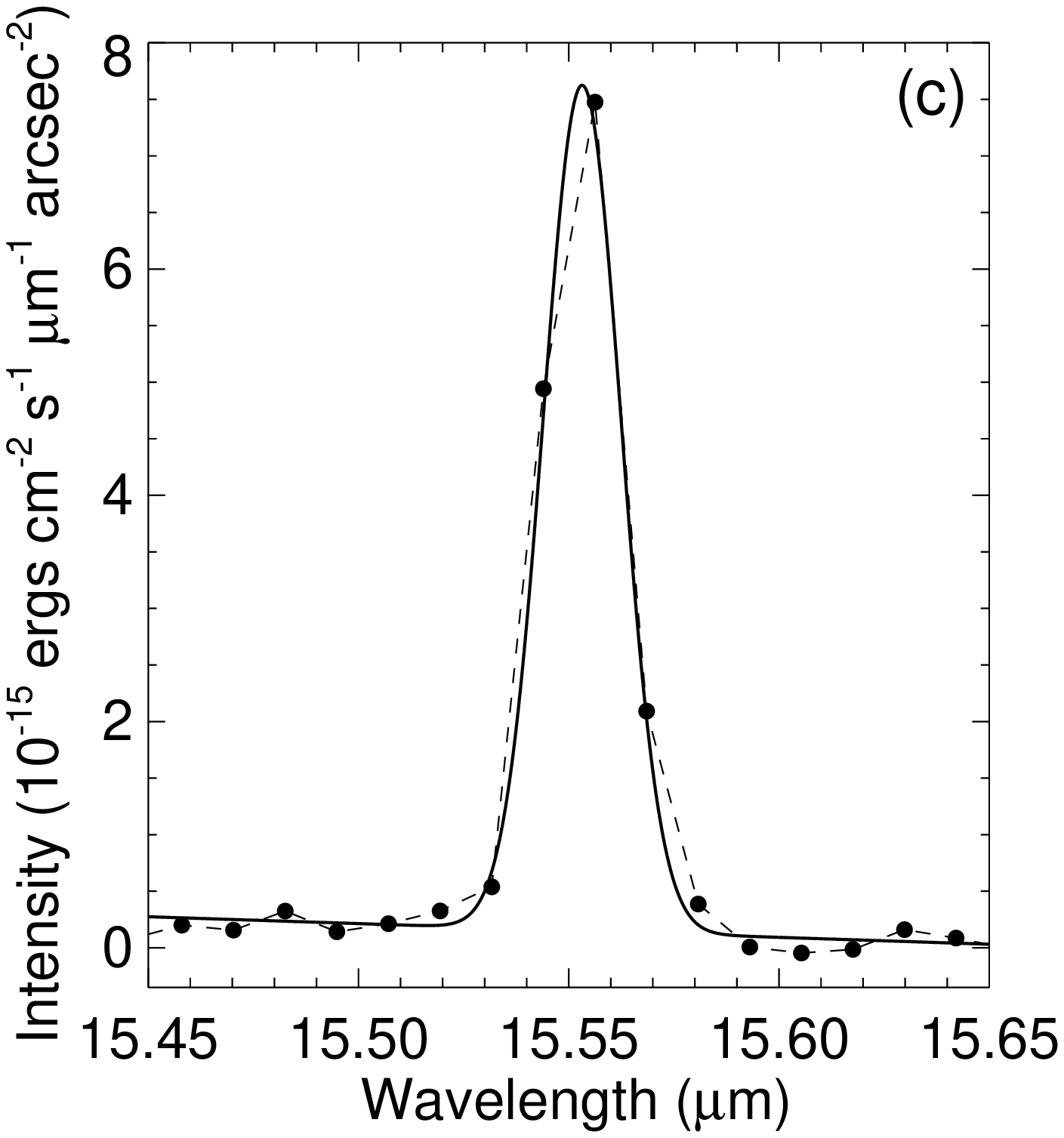}} &
\resizebox{7cm}{!}{\includegraphics{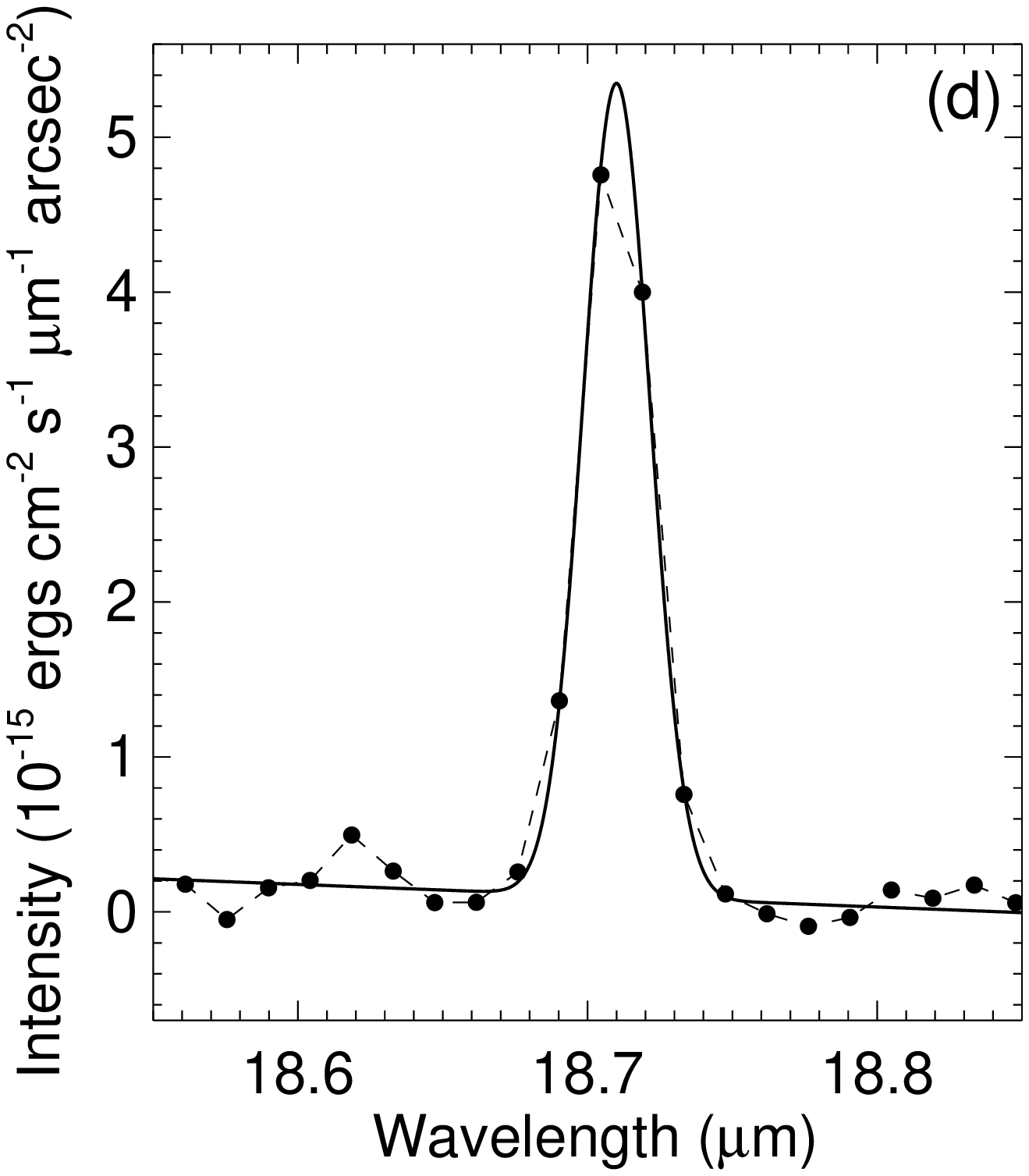}} \\
	\end{tabular}
    \caption[]{Measurements of the four emission lines in the H~{\sc ii} region KD~18 in
       NGC~6822: {\bf (a)} [S~{\sc iv}] 10.5;
       {\bf (b)} [Ne~{\sc ii}] 12.8;
       {\bf (c)} [Ne~{\sc iii}] 15.6; and
       {\bf (d)} [S~{\sc iii}] 18.7~$\mu$m.
       The data points are the filled circles and are connected by dotted lines.
       The fits to the continuum and Gaussian profiles are the solid lines.
       Such measurements provide the set of line fluxes for further analysis.}
\end{figure*}

In the Local Group dwarf irregular galaxy NGC 6822, 
we analyse {\it Spitzer} data for  9 H~{\sc ii} regions at a wide range of $R_{\rm G}$, from 0.067 to 1.37 kpc. 
The data were collected under the auspices of two programmes:
our  Cycle 4 programme 40910 and programme 193 (PI: R. Kennicutt)
from the {\it Spitzer} Legacy Programme `SINGS: The {\it Spitzer} Nearby Galaxies Survey' (Kennicutt et al. 2003). 
Our observations were made in 2008 from November 11 to November 29.    
These were observations of 
five H~{\sc ii} regions KD~2e, KD~5e, KD~10, KD~18, and KD~28e (Killen \& Dufour 1982). 
We obtained the data from the archives for the four others~-- Hu~I, Hu~III, Hu~V and Hu~X~-- 
all large sources known by their identification in Hubble (1925).
Most of those observations were made in 2004 between September 28 and 29 during campaign 750. 
An additional Astronomical Observation Request was made on October 10, 2005 during campaign 807.
For each of these 9 objects, there are IRS observations made with the
short-wavelength, moderate dispersion configuration 
called the short-high (SH) module (Houck et al. 2004). 
This configuration allowed for a spectral resolution of $\sim~600$ between
9.9 and 19.6~\micron\ 
and
provided cospatial data that cover all four of the emission lines [S~{\sc iv}] 10.51,
[Ne~{\sc ii}] 12.81,
[Ne~{\sc iii}] 15.56,
and [S~{\sc iii}] 18.71~$\mu$m
needed for our analyses.
For the brighter sources the hydrogen recombination line H(7-6) at 12.37~\micron\ 
was also observed.
The size of the SH aperture is 11.3 arcsec $\times$ 4.7 arcsec. 

Maps were arranged with the apertures overlapping along the
direction of the long slit axis (the `parallel' direction). 
In the direction of the short slit axis (the `perpendicular' direction), the
apertures were arranged immediately abutting each other; that is, with no overlap or space between them. 
We made 2-3 parallel steps at either 3.45 or 5.9 arcsec depending on the angular size
of the H~{\sc ii} region, and either 2 or 5 perpendicular steps whose size were either 2.3 or 2.65 arcsec.
This choice of aperture grid patterns (henceforth `{\it chex}', after the breakfast cereal) was used in
order to match the areas of the nebulae as closely as possible. The sizes of the chex were measured
using the {\it Spitzer} software {\sc spot}. The important feature of this grid overlapping is that most 
spatial positions will be covered in several locations on the array, minimizing the contribution 
of bad pixels to our raw data.	
Table 1 lists the H~{\sc ii} regions and the aperture grid configurations used to observe each.
{These locations are also shown in Fig.~1.}

To build our post-BCD (basic calibrated data) data products, we use 
{\sc cubism}, the CUbe Builder for IRS Spectral Mapping (version 1.70, Smith et al. 2007). 
{\sc cubism} was used to build spectral maps, account for our 
aperture grid overlaps, and to deal effectively with bad pixels. 
We used the `generate record bad pixels' function to automatically remove bad pixels 
that deviated by 5$\sigma$ from the median pixel intensity value in each BCD 
and occurred within at least 10~per~cent of the BCDs. 
Global bad pixels (those occurring at the same pixel in every BCD) were removed by hand. 
The result is a data cube, with two spatial dimensions and one spectral dimension. 

The {\it Spitzer} IRS calibration assumes that all objects are point sources. 
However, for NGC 6822, unlike M83 (R07) and M33 (R08), the H~{\sc ii} regions are extended 
compared to the IRS SH slits (for which reason each map covers more than a single slit aperture). 
Because of diffraction, 
a finite-sized aperture measures significant 
energy from outside the slit for an extended source compared to a point source
(or, a point source loses energy compared to an extended source). 
Corrections for this effect are called Slit Loss Correction Factors (SLCFs),
such that extended source intensities must be multiplied by the SLCFs in order to be properly calibrated. 
The {\sc cubism} option to build the data cubes multiplying by the SLCFs was used for NGC 6822, 
whereas it was not used for the much smaller (in arcsec) H~{\sc ii} regions in M33, 
also reduced with {\sc cubism} (R08).

\input table2.tex

The next step in our analysis of the spectra is the use of the line-fitting
routines of the IRS Spectroscopy Modelling Analysis and Reduction Tool ({\sc smart}; Higdon et al. 2004).
We used {\sc smart} to fit Gaussians to our measured emission lines, 
fitting the line simultaneously with the continuum, 
where the continuum baseline was fitted with a linear function. 
Figs 2(a)-(d) show the fits for each of the four lines in the H~{\sc ii} region KD~18.
The measured line intensities are given in Table~2. 
A line is deemed detected if the flux is above 3$\sigma$ uncertainty. 
Our uncertainties are calculated from the root-mean-squared deviation of the data from the fit; 
they do not include systematic effects.

In section 2 of our M83 (R07) and M33 (R08) papers, we discussed and also estimated
systematic uncertainties and how they affect the line fluxes. 
We repeat the major points of that topic here. 
SLCFs are probably the largest source of uncertainty
since, in fact, the observed H~{\sc ii}  regions are neither point sources nor truly extended.
The SLCF factors are 0.697, 0.663, 0.601 and 0.543 for the 10.5, 12.8, 15.6 and 18.7~\micron\ lines,
respectively. 
The factors were obtained by interpolating numbers provided by the $`$$b1\_slitloss\_convert.tbl$$'$
file from the {\it Spitzer} IRS Custom Extraction Tool ({\sc spice}) for the SH module. 
For the uniformly filled aperture the maximum uncertainty in the flux due to this effect 
would have been on the order of $\sim$40~per~cent for the [S~{\sc iii}] 18.7~\micron\ line.
In R07 and R08 we did not use SLCFs, and so in those papers we noted that the values for flux listed were
upper limits, and the application of the SLCFs would only work to lower them. 
Importantly, we also noted that our science depends on line intensity {\it ratios}, 
thus the possible uncertainty due to this effect in R08 would be lower, 
e.g. $\sim$ 28~per~cent when we deal with the line flux ratio [S~{\sc iv}~] 10.5/[S~{\sc iii}] 18.7~\micron.
Additional uncertainties, such as the absolute flux calibration and pointing error, 
are probably smaller and divide out in the computation of the line intensity ratios.
The systematic uncertainty far exceeds the measured uncertainty for all of our lines, even the faintest.

In addition to the line flux, the measured full width at half-maximum (FWHM) and heliocentric radial velocities ($V_{\rm helio}$) 
are listed in Table 2. 
Both the FWHM and $V_{\rm helio}$ serve as values by which to judge the measured line fluxes. 
With a resolving power of $\sim$600, our lines should have a FWHM of roughly 500~km~s$^{-1}$.
The values for $V_{\rm helio}$ 
should straddle the heliocentric systemic radial velocity for NGC 6822 of $-57$~km~s$^{-1}$
(Koribalski et al. 2004).
Most of our measurements are in agreement with these expectations.

\section{Results}

\input table3.tex

\subsection{Ionic abundances}

Ionic abundance ratios were computed with the standard semi-empirical analysis
(e.g., Rubin et al. 1994).
The references for the effective collision strengths are 
[S~{\sc iv}]: Saraph \& Storey (1999),
[Ne~{\sc ii}]: Griffin, Mitnik \& Badnell (2001),
[Ne~{\sc iii}]: McLaughlin et al. (2011),
and [S~{\sc iii}]: Grieve et al. (2014), with additional references in Simpson et~al.\  (2012).
Values of both electron temperatures ($T_e$) and densities ($N_e$) were assumed;
fortunately, at the mid-infrared (MIR) wavelengths of the {\it Spitzer} IRS SH module, the forbidden lines 
have very little sensitivity to either $T_e$ or $N_e$.
However, this is not the case for the H(7-6) line at 12.37 \micron, whose volume emissivity has 
a temperature sensitivity of $T_e^{-1.3}$ (emissivity values from Storey \& Hummer 1995).
Thus there could be some systematic bias in the abundances derived with respect to hydrogen
owing to mis-estimation of $T_e$.

A topic rarely considered is systematic uncertainties due to errors in the atomic data.
We remark here on the differences that are obtained as a result of updating 
the effective collision strengths.
For example, at $T_e = 8000$~K and $N_e = 100$ cm$^{-3}$,  the ratios of the volume emissivities  
$\epsilon$\ used in R07 and R08 to the current are 
[Ne~{\sc ii}] 12.8 \micron: $\epsilon$(Saraph \& Tully 1994)/$\epsilon$(Griffin et al. 2001) = 0.906,
[Ne~{\sc iii}] 15.6 \micron: $\epsilon$(McLaughlin \& Bell 2000)/$\epsilon$(McLaughlin et al. 2011) = 1.374,
and 
[S~{\sc iii}] 18.7 \micron: $\epsilon$(Tayal \& Gupta 1999)/$\epsilon$(Grieve et al. 2014) = 1.328.
(These ratios give the new derived abundances with respect to hydrogen 
divided by the previous.)
Another new set of cross sections for S$^{++}$ are by Mendoza \& Bautista (2014).
This set gives the ratio for 
[S~{\sc iii}] 18.7 \micron: $\epsilon$(Tayal \& Gupta 1999)/$\epsilon$(Mendoza \& Bautista 2014) = 1.073.
For lines like [S~{\sc iii}], an important test of the atomic data is whether the ratios 
of two observed lines agree with the theory in the low density limit, where the ratio
is dependent only on the collision strengths and not on the transition probabilities.
For the [S~{\sc iii}] 18.7/33.5 \micron\ line ratio, one must be careful to pick only low extinction 
sources; however, we observe that this ratio is lower than the theoretical low density limit for 
both the Grieve et al. (2014) and the Mendoza \& Bautista (2014) cross sections
(the latter worse than the former) in the outermost positions observed in M42
by R11 (see Table~A1 for updated densities) 
and in several low density, non-AGN galaxies observed by Dale et al. (2009).
We conclude that it is {\it crucial} that one uses the same atomic data when comparing abundances 
from multiple sources.
As far as the future goes, Mendoza \& Bautista (2014) remark that 
``in the past two decades ... [the] statistical consistency [of effective collision strengths] is around the $\sim$20--30\% level''.
Moreover, ``[t]his inherent dispersion ... would be hard to reduce in practice...''
Even so, the newer collision strengths should be better than the older ones because of upgrades to computer processing power and corresponding upgrades to computer codes.

{As stated above, we used the effective collision strengths of Saraph \& Storey (1999) 
for the [S {\sc iv}] 10.5 \micron\ line. 
We used this, rather than the collision strengths of Tayal (2000)
because those of Saraph \& Storey (1999) start at the low $T_e$ of 1000~K 
whereas those of Tayal (2000) are for 10,000~K or higher, and most of the H~{\sc ii} regions
that we observe have $T_e < 10,000$~K (for example, the sources in Appendices A1 -- A3).
Fortunately, the two sets of collision strengths differ by less than a per cent 
in the overlap region.
This year new collision strengths for [S {\sc iv}] have been published by 
Del Zanna \& Badnell (2016) for $T_e \ga 16,000$~K.
At that temperature they are only a few per cent larger than the collision strengths 
of either Saraph \& Storey (1999) or Tayal (2000).
We conclude that it is not likely that there is much uncertainty in the collision
strengths used for estimating the ionic abundance of S$^{3+}$ from the [S~{\sc iv}] 
measurements.}

\subsection{NGC 6822}

In Table 3 we present the ionic abundance ratios for the H~{\sc ii} regions in NGC 6822. 
In order to derive these ratios, our analysis assumed 
a constant $T_e$ (12,000~K) and $N_e$ (100~cm$^{-3}$)  
to obtain the volume emissivities for the four observed heavy element transitions plus H(7-6).
{$T_e = 12000$~K is the rounded average of values of $T_e$ from optical forbidden lines 
measured by Peimbert, Peimbert, \& Ruiz (2005); Lee, Skillman, \& Venn (2006); and Hern\'andez-Mart\'inez et al. (2009). 
$N_e = 100$~cm$^{-3}$ is an average of the densities found in Hu~V and Hu~X 
by Peimbert et al. (2005).}
For the ions Ne$^+$, Ne$^{++}$, S$^{++}$, and S$^{3+}$,
we use the  atomic data described in Simpson et~al. (2012) plus 
the updated effective collision strengths listed above.
For H$^+$ from the H(7-6) line, we used Storey \& Hummer (1995).
There is a bit of a complication here because at Spitzer's spectral
resolution, the H(7-6) line is blended with the H(11-8) line.
Their respective $\lambda$(vac) equal 12.371898 and 12.387168~$\mu$m.
In order to correct for the contribution of
the H(11-8) line, we used the relative intensity of H(11-8)/H(7-6)
from recombination theory (Storey \& Hummer 1995) assuming
$N_e$~= 100~cm$^{-3}$.
The  ratio H(11-8)/H(7-6)~= 0.122 and holds over a fairly wide range
in $N_e$ and $T_e$ 
appropriate for our objects.

We note again that there are two critical advantages compared with previous {\it optical}
studies of various other ionic ratios:
(1) While the collisionally-excited optical lines vary exponentially with electron temperature ($T_e$),
the IR {forbidden} lines have a weak yet similar $T_e$ dependence 
{(proportional to $\sim T_e^{-0.3}$)}, 
and (2) the IR lines suffer far less from interstellar extinction.
For our purposes,  the differential extinction correction
is negligible as the lines are relatively close in wavelength.
In our analysis, we are concerned with ionic abundance ratios
and therefore line flux ratios.
{The density dependence of the volume emissivities of the MIR lines is also small 
at the low densities of these extragalactic H~{\sc ii} regions since the critical densities 
are all greater than $10^4$ cm$^{-3}$ (Rubin 1989).
The major exception to the weak temperature dependence of the volume emissivities is the H 7-6 recombination line, as was mentioned in the previous section.}

Because NGC 6822 is a dwarf irregular galaxy, 
it does not have a disc-like structure such as the spirals M83 and M33.
Nonetheless we derived the deprojected $R_{\rm G}$ values for the 9 H~{\sc ii} regions,
ranging from 0.065 to 1.37 kpc. 
We found no statistically significant slope nor any trend in ionization 
(Ne$^{++}$/Ne$^+$ or S$^{3+}$/S$^{++}$) versus $R_{\rm G}$ for the NGC 6822 nebulae. 
Similarly, Lee {et al.} (2006) found no significant gradient 
in the oxygen abundance;
moreover, Hern\'andez-Mart\'inez et al. (2009) found that the central 2~kpc of NGC 6822  
is chemically homogeneous from their study of abundances in planetary nebulae and H~{\sc ii} regions.
We note, however, that Patrick et al. (2015) found a gradient of the metallicity of 
red super giants (but with low significance); 
these massive stars represent a young population similar to H~{\sc ii} regions. 

\section{Constraints on the Ionizing SED for the Stars Exciting the H~{II} Regions}


\begin{figure*}
    \begin{tabular}{cc}
    \resizebox{75mm}{!}{\includegraphics{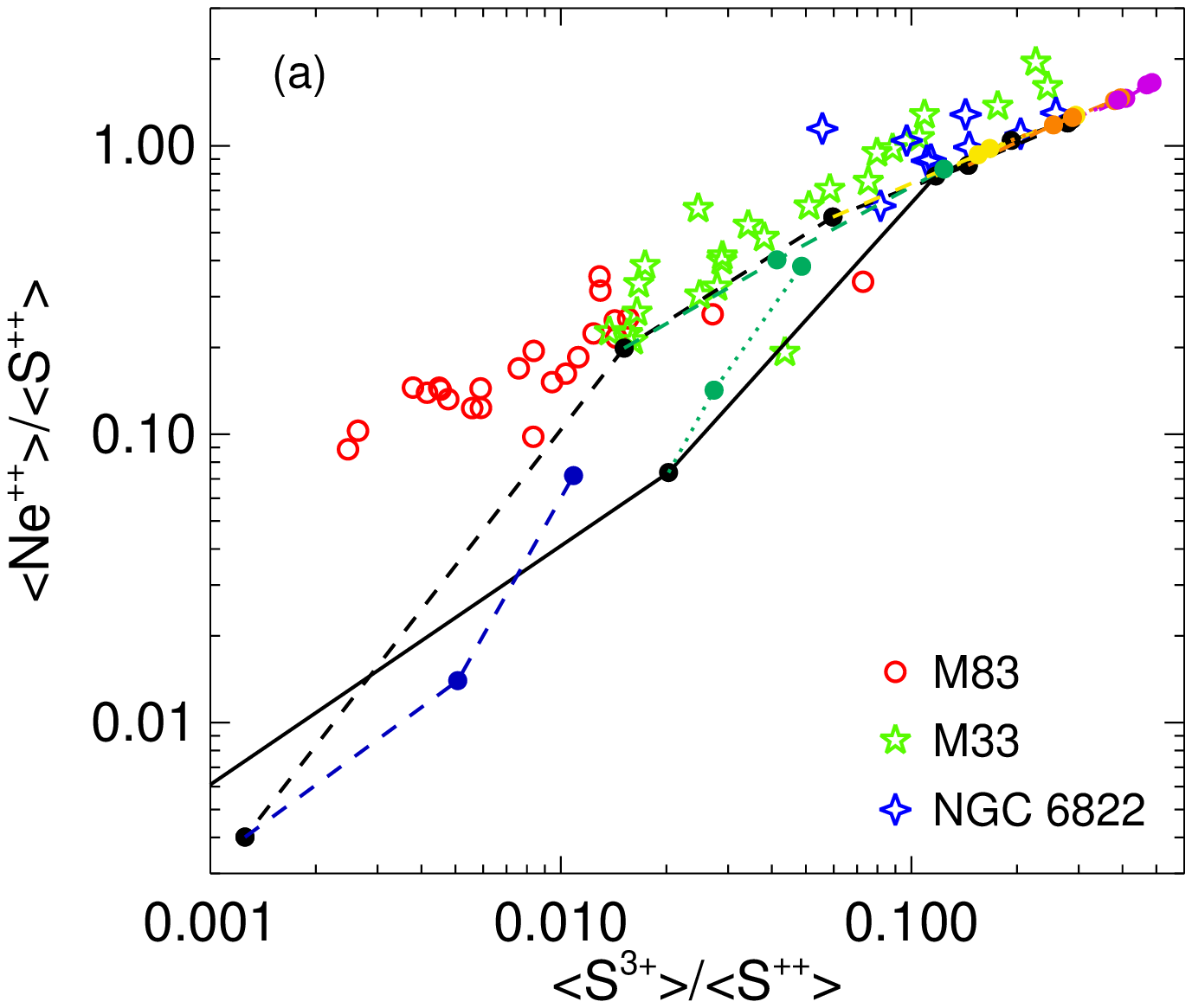}} &
     \resizebox{75mm}{!}{\includegraphics{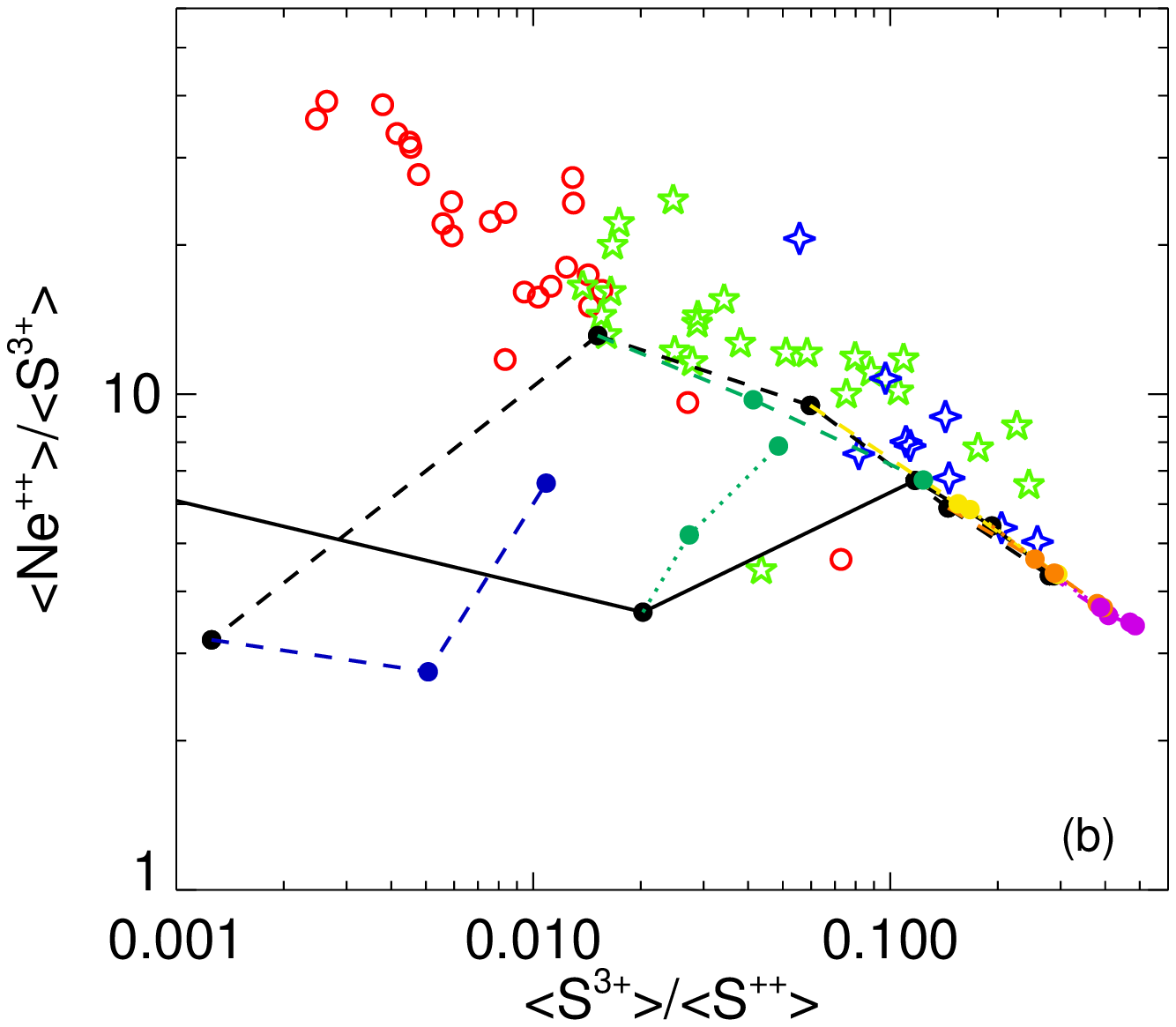}}
  \end{tabular}
     \resizebox{75mm}{!}{\includegraphics{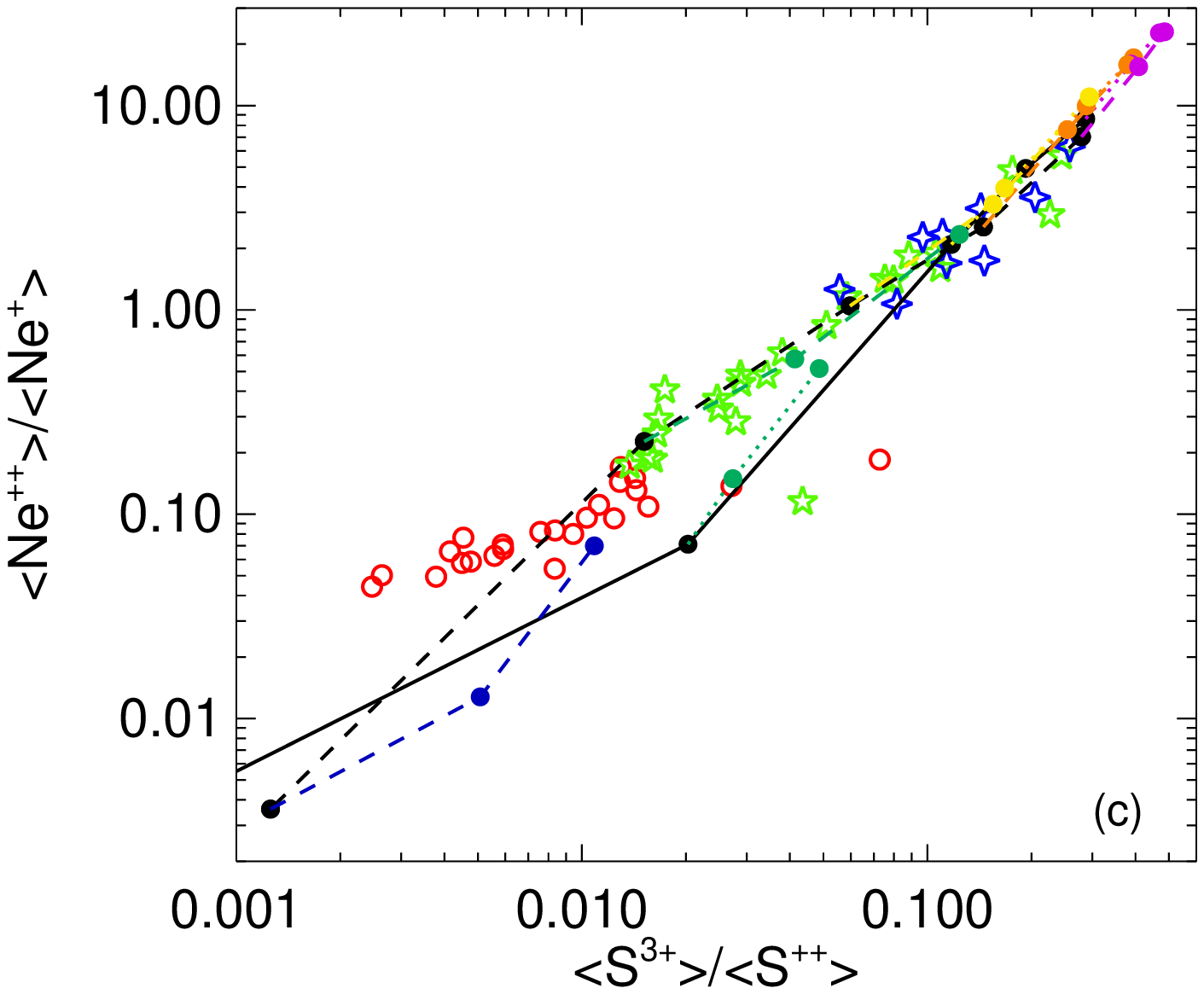}}
\caption[]{{\it {\bf (a)}}
Theoretical predictions of the fractional ionization ratios
$<$Ne$^{++}$$>$/$<$S$^{++}$$>$
versus 
$<$S$^{3+}$$>$/$<$S$^{++}$$>$,
computed using our photoionization code {\sc nebula}.
The lines connect the results of nebular models calculated with the
ionizing SEDs predicted from various stellar atmosphere models,
changing {\it no other parameter except the SED}.
For the H~{\sc ii} region models calculated with
Solar-abundance atmospheres,
{the solid line connects models with `dwarf' atmospheres and the dashed line
connects models with `supergiant' atmospheres (see text).}
Models computed with SEDs with 0.4 times solar abundance and 0.1 times solar abundance 
are connected to the 1.0 times solar abundance model with dashed and dotted lines 
for `supergiant' and `dwarf' atmosphere models, respectively.
Different values of \teff\ are represented by black dots and line color: 
\teff\ = 30~kK: blue; 
\teff\ = 35~kK: green; 
\teff\ = 40~kK: yellow; 
\teff\ = 45~kK: orange; and
\teff\ = 50~kK: purple. 
All the models with `dwarf' SEDs and \teff\ = 30~kK have 
($<$S$^{3+}$$>$/$<$S$^{++}$$>) \  < 0.001$ and so do not appear on the plot.
To compare our data with the models, we need to divide the observed
Ne$^{++}$/S$^{++}$
and
Ne$^{++}$/S$^{3+}$
ratios by an assumed Ne/S abundance ratio.
We use the Orion Nebula Ne/S~= 9.08 from Appendix A1.
The open red circles are our prior results for the M83 H~{\sc ii} regions (R07).
The green five-pointed stars are the M33 results from R08.
The blue four-pointed stars are our NGC 6822 results derived
from our line flux measurements (Table~3).  
{\it {\bf (b)}}
The same as panel (a) except
the ordinate is $<$Ne$^{++}$$>$/$<$S$^{3+}$$>$.
Both panels dramatically illustrate the sensitivity of the H~{\sc ii}
region model predictions of these ionic abundance ratios to the
ionizing SED that is input to nebular plasma simulations.
These data, for the most part, appear to track the `supergiant' locus.
{\it {\bf (c)}}
Similar to  panels (a) and (b) except
the ordinate is $<$Ne$^{++}$$>$/$<$Ne$^+$$>$.
}

\end{figure*}

In addition to being sensitive to the parameters of the H~{\sc ii} region gas, 
the various fractional ionic abundances found in H~{\sc ii} regions 
 are highly sensitive to the stellar ionizing SED. 
For NGC 6822, as for M83 (R07) and M33 (R08), we use our {\it Spitzer} data to investigate the
Ne$^+$ and Ne$^{++}$ fractional ionic abundances,
written as $<$Ne$^{+}$$>$ and $<$Ne$^{++}$$>$,
as well as those of  S$^{++}$ and S$^{3+}$ ($<$S$^{++}$$>$ and $<$S$^{3+}$$>$).
We demonstrate this with the ratios of fractional ionizations that are seen in Fig.~3:
$<$Ne$^{++}$$>$/$<$S$^{++}$$>$
versus 
$<$S$^{3+}$$>$/$<$S$^{++}$$>$
(Fig.~3a),
$<$Ne$^{++}$$>$/$<$S$^{3+}$$>$
versus 
$<$S$^{3+}$$>$/$<$S$^{++}$$>$
(Fig.~3b),
and Ne$^{++}$/Ne$^+$ versus 
$<$S$^{3+}$$>$/$<$S$^{++}$$>$
(Fig.~3c)
Our goals are
to test how closely our observed H~{\sc ii} regions are fit by these nebular models.
We calculate these ionic ratios using our photoionization code {\sc nebula}
(Rubin 1985; Simpson et~al.\  2004; Rodr\'\i guez \& Rubin 2005). 

In our studies of the parameters that affect the ionization structure of H~{\sc ii} regions,
we proceed by testing one parameter at a time.
In R07 and R08 we first tested the results of nebular models calculated with
SEDs predicted by various model stellar atmosphere codes,
changing nothing in the models except the SED. 
The model stellar atmospheres used in those papers 
represent five non-local thermodynamic equilibrium (non-LTE) models for massive O-stars
from the codes of Pauldrach et al. (2001), {Hillier \& Miller (1998) as computed by} 
Martins et al. (2005), and Lanz \& Hubeny (2003). 
In addition, we included the results from one set of LTE models (Kurucz 1992) 
because these LTE models have been used extensively in the past as input for H~{\sc ii} region models.
We concluded that from comparing non-LTE and LTE models, we
require a non-LTE treatment to arrive at more reliable SEDs for O-stars.

A second set of calculations in R07 and R08 tested a variety of nebular parameters 
with the atmosphere SED being the `dwarf' and `supergiant' models of Pauldrach et al. (2001),
{where the `dwarf' model atmospheres have log $g = 4.0$ 
and the `supergiant' model atmospheres have log $g = 3.0$ to 3.8, 
 where $g$ is the surface gravity in cm s$^{-2}$.}
The parameters tested were the gas nucleon density,
the total number of Lyman continuum photons ~s$^{-1}$ ($N_{\rm Lyc}$), 
the presence of a central cavity or a sphere with density extending up to the star, 
and nebular heavy element abundances. 
The results were that decreasing the abundances increased the ionization 
for a given stellar effective temperature, $T_{\rm eff}$ 
(increasing $<$Ne$^{++}$/Ne$^+>$ more than $<$S$^{3+}$$>$/$<$S$^{++}$$>$), 
and that changes that decreased the ionization parameter decreased the 
ionization (decreasing $<$S$^{3+}$$>$/$<$S$^{++}$$>$ more than $<$Ne$^{++}$/Ne$^+>$).
We refer the reader to R08 for a discussion of these and other possible effects.

In this paper we test the effect of changing the abundances in the ionizing star.
For these H~{\sc ii} region models we use a new grid of stellar atmosphere models 
(Weber, Pauldrach, \& Hoffmann 2015) 
computed with the {\sc wm-basic} code of Pauldrach et al. (2001) 
and Pauldrach, Vanbeveren, \& Hoffmann (2012).
This set of models uses the Solar abundances published in Asplund et al. (2009)
and also 0.4 and 0.1 times these abundances.
As in the previous compilation by Pauldrach et al. (2001), there are 
both {`dwarf' and `supergiant'} models.
Proceeding from the hot to the cool end,
the `dwarf' set has parameters 
($T_{\rm eff}$ in kK, log $g$ with $g$ in cm s$^{-2}$): 
(50, 4.0), (45, 3.9), (40, 3.75), (35, 3.8), and (30, 3.85)
while the `supergiant' set has
(50, 3.9), (45, 3.8), (40, 3.6), (35, 3.3), and (30, 3.0).

Figure 3 shows the ionization fractions computed with the new set of atmospheres. 
The nebular models calculated with the
`supergiant' stellar atmospheres are connected with dashed black lines. 
The solid black line connects models 
calculated with the `dwarf' stellar atmospheres. 

The nebular models in Fig. 3 are all of constant density,
ionization-bounded, and spherical. We used a constant nucleon
density  of 100~cm$^{-3}$ (as opposed to 1000~cm$^{-3}$ in R08)
that begins at the surface of the star. 
Each nebular model used a total number of Lyman continuum
photons~s$^{-1}$ ($N_{\rm Lyc}$)~= 10$^{50}$~s$^{-1}$.
For the nebular abundances, the ten elements in the code have their abundance by number relative to H as follows:
(He, C, N, O, Ne, Si, S, Ar, Fe) with
0.100, 3E$-$4 (Rubin et~al. 1993), 6.8E$-$5, 4.0E$-$4 (Rubin et~al. 1991), 
1.0E$-$4 (Rubin et~al. 2011), 4.5E$-$6 (Rubin et~al. 1993), 7.7E$-$6 (Rubin et~al. 2011), 
3E$-$6 (Simpson et~al. 1998), and 1.5E$-$6 (Esteban et~al. 2004; Rodr\'{\i}guez \& Rubin 2005), respectively.
This set of abundances represents the abundances of the Orion Nebula 
and is roughly a factor of 1.9 lower than the set we used in M83 (R07) and M33 (R08). 

The important thing to note in Fig. 3 is the increase in the excitation of the nebular models 
that accompanies the decrease in the abundances in the ionizing SEDs. 
Such an effect was suggested earlier by Mokiem et al. (2004), who computed  
stellar atmosphere models for {main sequence O-stars} using the  {\sc cmfgen} code 
(Hillier \& Miller 1998) with varying abundances. 
They found that a higher metallicity causes more opacity and softens the SED, 
and a lower metallicity does the opposite.

In R07 it was remarked that the nebular models appear to converge 
close to the higher ionization (hot \teff) side of Fig. 3
regardless of the values of \teff\ ($\sim$40--50~kK) and the different log~$g$
of the `dwarf' and `supergiant' models. 
Moreover, the nebular models computed with the 50,000~K atmospheres with 0.1 Z$_\odot$ abundances 
have almost twice the fractional ionizations of 
$<$Ne$^{++}$$>$/$<$Ne$^+$$>$ and $<$S$^{3+}$$>$/$<$S$^{++}$$>$.
A likely explanation for this convergent behavior in Fig. 3 is that 
at these high \teff\ values, the ionization balance in the atmospheres shifts
to higher ionization stages which have fewer lines capable of influencing the model calculations. 
Thus, blocking and blanketing effects no longer dominate the models as strongly, 
and the line radiation pressure is also less significant. 
This is even more noticeable for the low abundance atmospheres, 
where absorption by lines has even less effect.

Ratios of the ionization fractions for our galaxy data are also plotted in Fig. 3.
In order to compare our data with the models in Figs.~3(a),(b),
we need to divide the observed 
Ne$^{++}$/S$^{++}$ and Ne$^{++}$/S$^{3+}$
ratios by an assumed Ne/S abundance ratio,
taking into account the missing S$^+$ and S$^{4+}$.
For this, we adopt a constant Ne/S~= 9.08~$\pm$~0.16, the Orion Nebula value  
(R11) as updated in the Appendix.
The open red circles are our results for the M83 H~{\sc ii} regions.
The green 5-pointed stars are the M33 results 
and the blue 4-pointed stars are the results for the NGC 6822 nebulae, 
both adjusted by the assumed Ne/S and
derived from our observed line fluxes
using $N_e$ of 100~cm$^{-3}$ (Table 3).
The trends of the ionic ratios established from the prior M83 and M33 studies
are remarkably similar, and continue to higher ionization with the
present NGC 6822 objects.
There are two sources, one in M33 and one in M83,
that are deviantly low compared with
the theoretical tracks and the other data; these are suspect.
For M33, it is BCLMP~702.
The deviant object in M83
(in all three panels) is source RK~268.

One final comment: We notice here that the estimated fractional ionizations of 
 $<$Ne$^{++}$$>$/$<$S$^{++}$$>$ and  $<$Ne$^{++}$$>$/$<$S$^{3+}$$>$
in Figs 3a and 3b, respectively, mostly lie above all the model fractional ionizations.
We consider whether this may be the result of using too low a value for 
the Orion Nebula Ne/S normalization factor, Ne/S~= 9.08 (Appendix A1).
For these M42 data, the fractional ionization of S$^+$ ranges from 0.127 to 0.148 
for the eight M42 positions used for the average Ne/S.
However, for the models plotted in Fig. 3, the fractional ionization of S$^+ \sim 0.05$
for all the models except the `dwarf' \teff\ = 30,000 K models (all of which have too 
low $<$S$^{3+}$$>$/$<$S$^{++}$$>$ to fall on the plot).
The other missing ion is S$^{4+}$, which can have fractional ionizations of 
a few per cent in the models ionized by the hottest stars.
Such corrections could decrease the estimated fractional ionizations of 
 $<$Ne$^{++}$$>$/$<$S$^{++}$$>$ and  $<$Ne$^{++}$$>$/$<$S$^{3+}$$>$
by up to $\sim 20$ per cent, improving the fit of the data to the models.

\section{Discussion}

\subsection{NGC 6822}

Recently, Garc\'ia-Rojas et al. (2016) estimated the abundances of neon and sulphur 
from previous optical observations of Hu~V, Hu~X, and KD~28e, 
which they refer to as H{\sc ii}~15.
Our measurements of the S$^{++}$/H$^+$ ratio in Hu~V and Hu~X are 10--20 per cent lower 
than those of Garc\'ia-Rojas et al. (2016), within the uncertainties of the measurements,
but our measured Ne$^{++}$/S$^{++}$ is higher for all three H~{\sc ii} regions,
still within the uncertainties of the measurements.
These differences are probably due to (1) the completely different set of lines 
used in the measurements and (2) the completely different set of collision strengths 
used to estimate the ionic abundances. 
It is nice to see such good agreement -- we infer that other potential sources of error 
such as electron temperature or extinction are being handled adequately.

However, there is not such good agreement with the total abundance ratios S/H 
and Ne/S, although the Ne/H ratio is within the measurement uncertainties. 
Both optical and infrared observations of total abundances in ionized nebulae 
have to correct for unseen ionization states, 
here Ne$^+$ and S$^{3+}$ for optical observations, 
S$^+$ for infrared observations (the infrared is fortunate in being able 
to observe all of the ionization states of neon seen in H~{\sc ii} regions),
and S$^{4+}$ for both.
Corrections for missing ionization states are usually handled by reference to 
models; these are known as ionization correction factors (ICFs).
In our H~{\sc ii} region models the fraction of neutral neon is always less 
than the fraction of neutral hydrogen  
and the fraction of S in S$^{4+}$ is an order of magnitude or more lower than 
the fraction of S in S$^{3+}$, so both ions are rightfully ignored (this would not  
be correct for planetary nebulae).
The fraction of S in S$^+$ is inversely a function of the ionization parameter 
(the diluteness of the radiation field, Simpson et al. 2007);
it is not a strong function of the effective temperatures of the exciting stars.
Garc\'ia-Rojas et al. (2016) found ratios S$^+$/S$^{++}$ of 0.056, 0.14, and 0.087 
for Hu~V, KD~28e, and Hu~X, respectively.
This shows that the fraction of S in S$^+$ is a small but non-negligible contribution 
to the sulphur abundance, as we discussed above for M42 and the H~{\sc ii} region models.
Likewise, our analysis of our [S~{\sc iv}] measurements shows that S$^{3+}$ 
is also a small but non-negligible contribution to the sulphur abundance (Table~3, Fig.~2).
This is in contradiction to the very large contribution from unseen ionization 
states of sulphur (the sulphur ICF) that Garc\'ia-Rojas et al. (2016)
must have used to get their total S/H abundance.
We recommend that optical observers consider the now extensive literature 
of infrared measurements of neon and sulphur in H~{\sc ii} regions 
and planetary nebulae when devising ICFs for these elements.

\subsection{Neon to sulphur abundance ratio}


\begin{figure*}
\vskip-0.3truein
\centering
\resizebox{14.0cm}{!}{\includegraphics{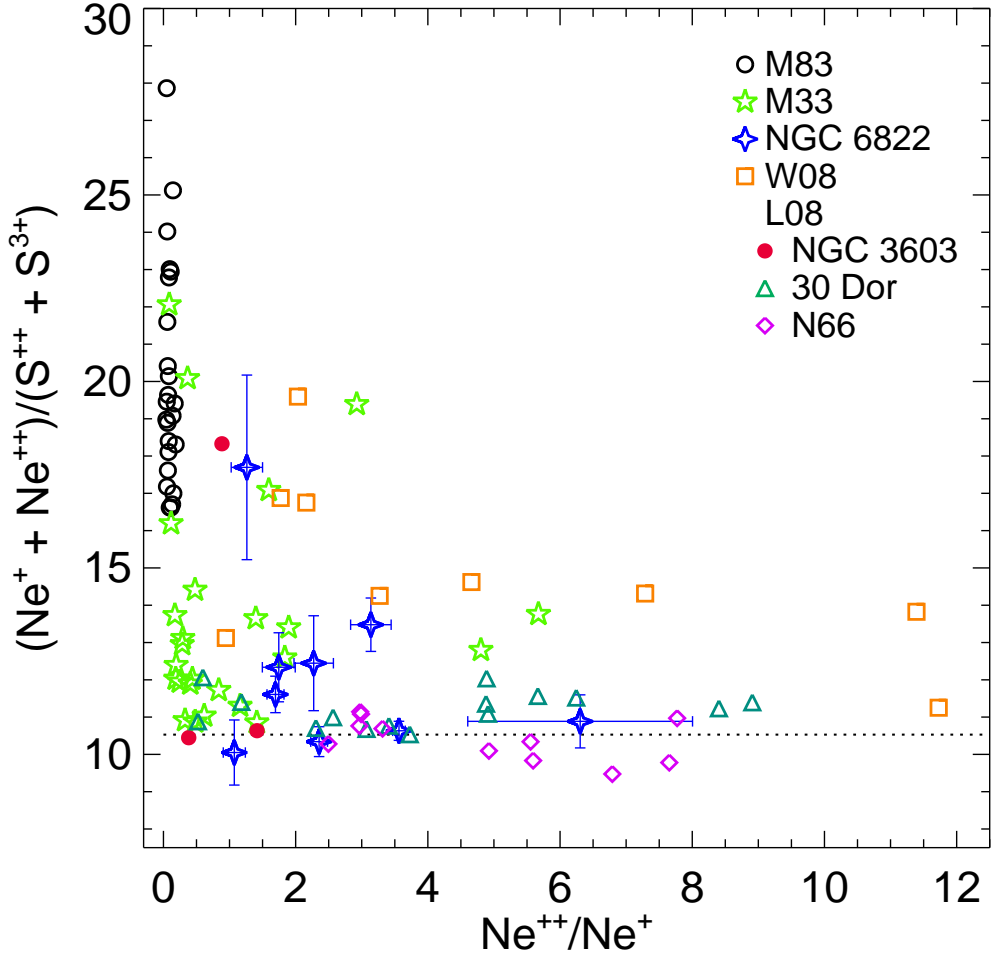} }
\caption[]{Plot of (Ne$^+$ + Ne$^{++}$)/(S$^{++}$ + S$^{++}$) versus Ne$^{++}$/Ne$^+$.
Here we show our results as blue 4-pointed stars for the 9 H~{\sc ii} regions in NGC~6822.
Our recomputed M33 results are shown as green 5-pointed stars
and our recomputed M83 results are shown as black circles.
These data demonstrate a huge variation in the inferred Ne/S
ratio when Ne$^{++}$/Ne$^+$ is low.
The orange squares show the Wu~et~al.\ (2008) data for
blue compact dwarf galaxies, as reanalyzed in Table A4.
We show only 9 points, those objects where they actually detected {\it all
four} lines: [S~{\sc iv}], [Ne~{\sc ii}], [Ne~{\sc iii}], and [S~{\sc iii}].
The median (average) Ne/S for the 9 galaxies is 14.3 (15.0).
The Orion value of 10.5 is shown as the dotted line.
The Lebouteiller~et~al.\ (2008) data were also reanalyzed here (Table A5) and are
presented as follows: NGC~3603 (red dots), 30~Dor (dark green triangles),
and N66 (purple diamonds).
The median Ne/S ratios for each are 10.6, 11.2, and 10.3, respectively.
}
\end{figure*}


\begin{figure*}
\resizebox{14.0cm}{!}{\includegraphics{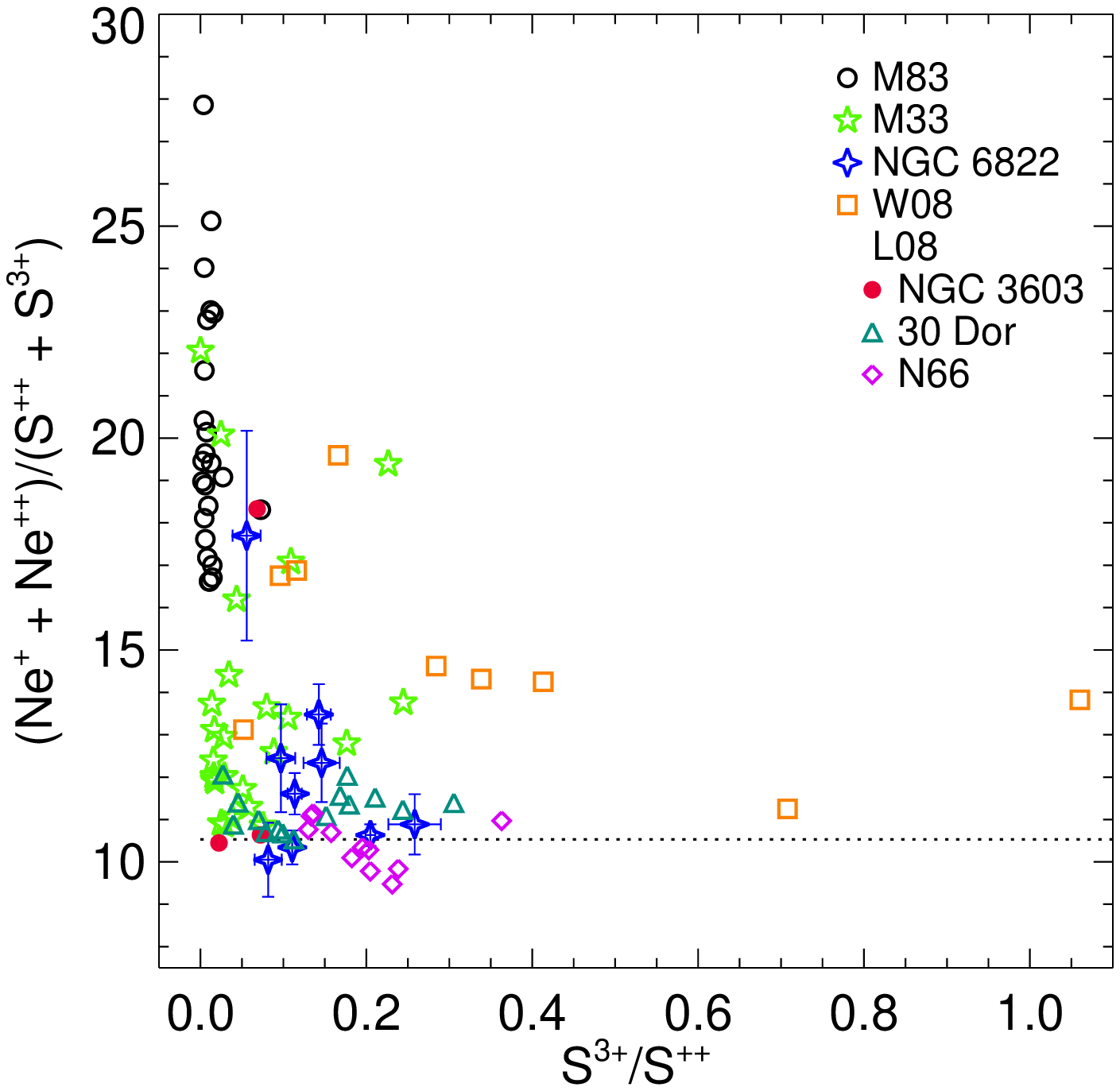} }
\vskip0.1truein
\caption[]{Plot of (Ne$^+$ + Ne$^{++}$)/(S$^{++}$ + S$^{++}$) versus S$^{3+}$/S$^{++}$.
The data show a huge variation in the inferred Ne/S
ratio at low ionization, when S$^{3+}$/S$^{++}$ is small.
The dotted line gives the Orion Nebula value of Ne/S$_{\rm MIR}$ = 10.5.
}
\end{figure*}

As in our previous work, in this section we consider only our MIR observations
of Ne and S and approximate Ne/S as Ne/S$_{\rm MIR}$ = (Ne$^+$ + Ne$^{++}$)/(S$^{++}$ + S$^{3+}$), 
the dominant ionization states of Ne and S in H~{\sc ii} regions. 
As stated above, in using this approximation we neglect contributions from S$^+$ and S$^{4+}$,
which we expect to be present to some degree. 
Otherwise, we have safely ignored neutral S and Ne since we are studying ionized regions. 

In R07, our Ne/S$_{\rm MIR}$ ratio in M83 had a statistically significant drop with increasing $R_{\rm G}$. 
In that paper we argued that this slope was not representative of a true gradient in Ne/S, 
but rather was due to not accounting for sulphur in other forms: S$^+$, molecular gas, and dust. 
There we postulated that an increasing fraction of S$^+$ with smaller $R_{\rm G}$ results 
in a constant Ne/S value throughout the galaxy. 
In M33 (and to a smaller extent in M83), 
there is an increasing degree of ionization with increasing $R_{\rm G}$. 
As mentioned in R08, M33 had a steeper positive ionization gradient than M83 
and yet our Ne/S$_{\rm MIR}$ estimates for M33 showed no correlation with $R_{\rm G}$. 
This at first seems suspect; however, the sulphur (but not neon) may be tied up in dust grains and molecular gas. 
For galaxies with higher metallicity (and hence higher dust content),
we expect more S to be included in that dust, 
rather than in any ionized state that we could detect with the IRS. 
We suspect that this is a likely reason for an elevated Ne/S$_{\rm MIR}$ detection in high metallicity H~{\sc ii} regions.
We detected no significant trend in Ne/S versus $R_{\rm G}$ in NGC 6822. 
It should be noted that
NGC 6822 is to a high degree chemically homogeneous with $R_{\rm G}$ (Hern\'andez-Mart\'inez et al. 2009),
and so a flat Ne/S versus $R_{\rm G}$ relation is expected. 
At 0.3Z$_{\odot}$ (Galametz et al. 2010), NGC 6822 has lower metallicity than M83 and M33, 
and it also happens to have a lower mean Ne/S$_{\rm MIR}$ value than those two objects. 
With the inclusion of sulphur but not neon into dust and molecular gas, a higher metallicity galaxy
will appear to have relatively less sulphur and therefore a higher Ne/S value. 
We conclude that the derived Ne/S$_{\rm MIR}$ ratios in low metallicity galaxies are representative of reliable Ne/S values. 

The Ne/S$_{\rm MIR}$ abundance ratios that we derived for 24 H~{\sc ii} regions in M83 varied from 27.9 to 16.6
(R07 with corrections for updated cross sections, see Table~A2 in the Appendix). 
All of these values are significantly higher than the Orion Nebula value of Ne/S = 9.08 
[{the average} (Ne$^+$ + Ne$^{++}$)/(S$^{++}$ + S$^{3+}$) = {10.5, see Appendix A1}]
and many are greater than any of those in M33 or NGC 6822. 
As stated in R08, we are considering our M83 Ne/S$_{\rm MIR}$ values as upper limits, 
while those for M33 and certainly those for NGC 6822 are likely robust estimates of a true Ne/S ratio.

As was mentioned before, because NGC 6822 and M33 H~{\sc ii} regions are of lower
metallicity and significantly higher ionization than those in M83, the amount of any correction
needed for S in forms other than S$^{++}$ and S$^{3+}$ is much lower. 
While they should still be considered upper limits, 
these M33 and NGC 6822 values are a much better estimate 
of a true Ne/S ratio than those derived for M83. 
The median (average) Ne/S$_{\rm MIR}$ ratio derived for H~{\sc ii} regions in M33 
is 12.8 (13.7$\pm$0.6). 
The median (average) Ne/S$_{\rm MIR}$ ratio derived for the 9 H~{\sc ii} regions in
NGC 6822 is 11.6 (12.2$\pm$0.8).
{Both medians and averages are given -- medians control for the presence 
of outliers such as a source with excitation parameters very remote from the 
rest of the group whereas averages give better statistics on the variance 
or spread of the data.}
The errors {associated with each average} are standard deviations of the mean, 
to apply to the galaxies as a whole.

\subsection{Comparison to other galaxies}

\input table4.tex

The reliability of nucleosynthesis and GCE models depends on how rigorously tested they are and
how their parameters are constrained. 
We provide fundamental observational data, which are pivotal in
the determination of the reliability of these models. 
A particularly valuable supplement relating to these pursuits would be to find out 
how much the Ne/S ratio can vary or whether or not there is a fairly universal value. 

Including the measurements of W08, L08, 
R07, and R08, we are in a position to further examine this Ne/S ratio. 
From their {\it Spitzer} observations of 13 blue compact dwarf galaxies, W08 found 
an average Ne/S$_{\rm MIR} = 12.5 \pm 3.1$. 
In that sample of galaxies, they found no correlation between the Ne/S ratio
and metallicity (Ne/H ratios). 
Just as we did with NGC 6822, we employ the same semi-empirical analysis code 
to their table 3 of line fluxes to derive the various ionic abundance ratios. 
Our new results are in Table A4 of the Appendix.
For these blue compact dwarf galaxies, we obtain a somewhat higher Ne/S$_{\rm MIR}$ ratio median (average) of 14.3 ($15.0 \pm 2.4$). 
As mentioned in R08, this was obtained by deriving the Ne$^{++}$/Ne$^+$ and S$^{3+}$/S$^{++}$ ratios for all their galaxies except 4 with at least one upper limit on a line flux.

In Fig. 4, we plot (Ne$^+$ + Ne$^{++}$)/(S$^{++}$ + S$^{3+}$) versus Ne$^{++}$/Ne$^+$
for our NGC 6822 results (blue 4-pointed star symbol) using the same
9 sources listed in Table 1. We plot these against our
Orion standard (Ne$^+$ + Ne$^{++}$)/(S$^{++}$ + S$^{3+}$) value of 10.5, which is shown as the horizontal dotted line. 
The results of our prior M83 study (R07) are shown as black
circles and those from the M33 study (R08) are shown as green 
stars. The wide spread in the M83 points dramatically demonstrates
the limitations of our method to infer Ne/S (except as an upper
limit) when Ne$^{++}$/Ne$^+$ is low (i.e., for low ionization objects).
The orange squares show the W08 data for the nine blue compact
dwarf galaxies, as discussed above.

Fig. 4 also shows our reanalyzed results of L08 (see Table A5). 
The three positions in NGC 3603 are plotted as red dots, the 15 positions in 30 Dor 
as dark green triangles, 
and the 9 positions in N66 are plotted as purple diamonds. 
The mean metallicities of the six galaxies and H~{\sc ii} regions
from ours and L08's papers are given in Table 4, 
where Ne/S$_{\rm MIR}$ ranges from $\sim$10.4 to $\sim$13.7. 

A likely explanation for the distribution of data points in Fig. 4 is that at
the lowest ionizations there is still a high density of sulphur present in the form of S$^+$  
and in molecular gas and dust. 
The three 30 Dor positions with the lowest ionization, as measured by both Ne$^{++}$/Ne$^+$ and 
S$^{3+}$/S$^{++}$ are 10, 17, and 8 in Table A5. Upon inspection of fig. 2 in L08, the positions of
\#10 and \#17 are near the periphery of their provided 30 Dor image,
and so they are expected to be characterized by lower ionization.
What we expect is that once all other species of S
become negligible with respect to S$^{++}$ and S$^{3+}$,
a horizontally asymptotic Ne/S will be approached at the higher ionizations.

	It is tantalizing to conjecture that for high-ionization 
objects, all the dominant states of Ne and S are measured, 
resulting in a robust estimate of the true Ne/S abundance ratio.
For objects as diverse as the Orion Nebula, NGC 3603, 30 Dor,
N66, the NGC 6822 H~{\sc ii} regions, and the blue compact dwarf
galaxies, there is remarkably little variation in the Ne/S derived.
All of these Ne/S values point to a much larger ratio than the `canonical'
Solar value of $\sim 6.5$ (Asplund et al. 2009) 
and what had previously been predicted by the GCE  models mentioned in Section 1. 
The question of how universal the Ne/S ratio may be will still require
further study. It is possible that it does take such a high degree
of ionization to percolate off any substantial amount of S that may
still be tied up in grains, molecules and S$^+$. {\it Spitzer}
observations of NGC 6822 have been useful in helping to determine
whether this is a valid presumption, and additional {\it Spitzer}
observations of high ionization objects will further elucidate the probability
of a universal Ne/S ratio. However, for still higher ionization sources,
such as planetary nebulae, 
the same type of analysis {as is} done here will not hold, because substantial amounts of Ne
will exist as Ne$^{3+}$ or higher and/or S as S$^{4+}$ or higher.

\subsection{Relationship between the Ne/S ratio and metallicity}	

In order to more closely examine a possible relationship between Ne/S abundances and metallicity
we calculated the mean {and median} 
Ne/S$_{\rm MIR}$ values from our sample H~{\sc ii} regions for two of our galaxies, M33
and NGC 6822. 
In Table~4 we combine these values with those {recalculated }
for the LMC H~{\sc ii} region 30 Dor, SMC emission nebula NGC 346 (N66), 
and the massive Galactic emission nebula NGC 3603 {in the Appendix}. 
In addition, we include our revised Ne/S$_{\rm MIR}$ value for M42,
which can be compared to the recent optical VLT observations 
of the Orion nebula protoplanetary disk LV 2, 
where the Ne/S abundance ratio was found to be 13.5 in the stellar jet (Tsamis \& Walsh 2011).
{Table~4 also contains estimated values of }
Z/Z$_{\odot}$ and the oxygen abundance (12 + log(O/H)). For sources with Z/Z$_{\odot}$ values not referenced, we calculated metallicity
based on comparing their (12 + log(O/H)) value with the solar value 8.69 $\pm$ 0.05 (Asplund et al. 2009).  For those sources with the oxygen abundance
not referenced, we provide a value based on the ratio of Z/Z$_{\odot}$ of that source and the solar oxygen abundance.

{We consider whether the observed Ne/S ratio could be a function of metallicity. 
We find no such correlation in the sources/galaxies of Table~4, although none 
of them is a truly high-metallicity source like the H~{\sc ii} regions of M83.
High-metallicity objects tend to be low excitation, with more of the sulphur 
existing as S$^+$, thereby increasing the apparent Ne/S if it is computed 
only from (Ne$^+$ + Ne$^{++}$)/(S$^{++}$ + S$^{3+}$), as is seen in Figs. 4 and 5.
Both high- and low-metallicity sources will be required to get enough 
of a baseline for good tests of Ne/S versus metallicity.
In the future, extragalactic H~{\sc ii} regions will be observed extensively
with MIRI on {\it JWST} in coordination with optical studies of the S$^+$ abundance.
It probably will require observations of high-metallicity objects to 
determine whether sulphur can be depleted onto dust grains as well as molecular gas,
something of which we currently have little knowledge.}



\section {Summary and conclusions}

We have observed emission lines of [Ne~{\sc ii}] 12.81,
[Ne~{\sc iii}] 15.56,
[S~{\sc iii}] 18.71, and [S~{\sc iv}] 10.51~$\mu$m
cospatially with the {\it  Spitzer Space Telescope} using the Infrared
Spectrograph in short-high mode.
From the measured fluxes, we estimate the ionic abundance ratios 
Ne$^{++}$/Ne$^+$,
S$^{3+}$/S$^{++}$, and S$^{++}$/Ne$^+$ in nine H~{\sc ii} regions in the
local dwarf irregular galaxy NGC~6822. By sampling the dominant
ionization states of Ne and S for H~{\sc ii} regions, we can approximate
the Ne/S ratio by (Ne$^+$ + Ne$^{++}$)/(S$^{++}$ + S$^{3+}$). For NGC 6822,
we have found no significant variation in the Ne/S ratio with $R_{\rm G}$.
Both Ne and S are products of $\alpha$-chain reactions following
carbon and oxygen burning in stars, with large production factors
from core-collapse supernovae. Both are primary elements, making
their yields depend very little on stellar metallicity. Thus,
at least to a `first order', it is expected that Ne/S remains
relatively constant throughout a galaxy. As discussed in Sections 3
and 5, our estimate for Ne/S has accounted for neither the presence 
of S$^+$ nor S that may be tied up in grains or molecular gas.

These NGC 6822 data, combined with our previous {\it Spitzer} data
(see Figs. 4 and 5), are consistent with our view that there are now
reliable estimates for the {\it total} Ne/S ratio. As long as the level
of ionization is sufficiently high so that the amount of sulphur in forms
other than S$^{++}$ and S$^{3+}$ is small, the methodology used here will
provide a robust total Ne/S estimate. 
The median Ne/S value we derive for NGC 6822 is 11.6. 
For the W08 blue compact dwarf galaxies in Table A4, we find a 
median Ne/S for the nine galaxies (and also the five higher ionization galaxies) of 14.3. 
From Table~A5 with the recomputed Ne/S from the L08 observations, 
we find a  median Ne/S of 10.3 for N66 and 11.2 for 30 Dor. 
N66 has the lowest metallicity ($\sim$0.2 solar)
while 30 Dor in the LMC has an intermediate value compared with the Milky Way.
{We find no correlation of Ne/S with metallicity.
These galactic values of Ne/S are substantially higher than the Ne/S = 6.5 
estimated for the Sun by Asplund et al. (2009). 
We direct the reader to the Summary and Conclusions section of R08 for further discussion.}

The data set here, combined with our previous M83 (R07) and M33 (R08) results
as reanalysed in the Appendix,
may be used as constraints on the ionizing SEDs for the stars exciting these nebulae. 
To do this we 
compared the ratio of fractional ionizations $<$Ne$^{++}$$>$/$<$S$^{++}$$>$,
$<$Ne$^{++}$$>$/$<$S$^{3+}$$>$,
and
$<$Ne$^{++}$$>$/$<$Ne$^+$$>$
versus 
$<$S$^{3+}$$>$/$<$S$^{++}$$>$
with predictions made from our photoionization models.
In this comparison we assumed that the Ne/S ratio does not vary and equals the Orion Nebula value. 
Generally,
the best fit is to the nebular models using the `supergiant' stellar atmosphere
models (Pauldrach et al. 2001; Pauldrach et al. 2012; Weber et al. 2015) 
computed with the {\sc wm-basic} code. The comparison shown
in Fig. 3(c) is independent of the Ne/S ratio. 
The effect of decreasing the metallicity in the stellar atmosphere models 
is to increase the ionization state of the H~{\sc ii} region models
because the ionizing SED is harder. 
However, these results are mainly qualitative because
there is also sensitivity to the nebular parameters
such as density and chemical composition, as was shown in R08.

\section*{Acknowledgements}

This work is based on observations made with the {\it Spitzer Space Telescope},
which is operated by the Jet Propulsion Laboratory (JPL),
California Institute of Technology under a contract with NASA. Support for
this work was provided by NASA for this {\it Spitzer} programme, 40910,
and by the Deutsche Forschungsgemeinschaft (DFG) under grant PA 477/19-1. 
The IRS was a collaborative venture between Cornell University
and Ball Aerospace Corporation 
funded by NASA through the JPL and Ames Research Center.
{\sc smart} was developed by the IRS Team at Cornell University and is available
through the Spitzer Science Center at Caltech.
{This research has made use of the NASA/IPAC Extragalactic Database (NED), which is operated by the Jet Propulsion Laboratory, California Institute of Technology, under contract with the National Aeronautics and Space Administration.}
We thank Ted Kietzman, Spencer Ledoux, Divya Ramakrishnan, Vikram Sivaraja, and Scott Zhuge for their assistance with the data reduction
{and we thank the referee for the thoughtful comments that greatly improved the presentation of this paper.}

\begin{appendix}
\appendix

\section{Updated abundance calculations}

Because we have updated the atomic data for neon and sulphur in our 
ionic abundance calculation program, we here 
repeat the results for M42 (R11), 
M83 (R07), and M33 plus comparison H~{\sc ii} regions (R08) as they are affected by these revisions.
For each H~{\sc ii} region we assumed an electron temperature and 
computed the ionic abundances as described earlier in Section 3.1.

\subsection{M42}

\input tableA1.tex

The recomputed ionic abundance ratios for M42 are given in Table A1.
Because the Orion Nebula (M42) is often used as the standard 
for abundances in the interstellar medium, 
we re-estimate the total abundances Ne/H and S/H using the observations described in R11.
Since the longer wavelength [S~{\sc iii}] 33.5 \micron\ lines were also observed 
in M42 with the {\it Spitzer} IRS long-high (LH) module 
(being very careful to cover the SH and LH apertures on the sky identically), 
we first computed the electron densities from the [S~{\sc iii}] 18.7/33.5 \micron\ line ratios,
as described in R11.
To estimate the abundance of S$^+$, R11 also took their optical spectra for each location that 
was closest to the {\it Spitzer} spectrum location (`chex') and 
determined the S$^+$/S$^{++}$ ratios from the [S~{\sc ii}] 0.6716 and 0.6731 \micron\ lines 
and the [S~{\sc iii}] 0.6312 \micron\ lines.
We repeated this computation using the updated atomic data for S$^+$ 
of Mendoza \& Bautista (2014), $T_e = 8000$~K, and $N_e$ computed from 
the [S~{\sc ii}] 0.6716/0.6733 line ratios.
Then for each of the eight inner and middle positions, which have the highest ionization,
we estimated the abundance of S$^+$/H$^+$ from the S$^{++}$/H$^+$ ratio
observed with {\it Spitzer} (Table A1)  
and computed the (Ne$^+$ + Ne$^{++}$)/(S$^+$ + S$^{++}$ + S$^{++}$) ratios.
The mean and standard deviation of the mean for these eight positions is Ne/S $= 9.08 \pm 0.16$.
With no corrections for S$^+$, the (Ne$^+$ + Ne$^{++}$)/(S$^{++}$ + S$^{++}$) ratio
is $= 10.5 \pm 0.2$.
With respect to hydrogen we have Ne/H $= 1.02 \pm 0.03 \times 10^{-4}$
and corrected for $S^+$, S/H $= 11.1 \pm 0.4 \times 10^{-6}$.
The lower limits to $N_e$ for the last two positions are due to their having 
their [S~{\sc iii}] 18.7/33.5 \micron\ line ratios less than the theoretical ratio 
for any density.

\subsection{M83}

\input tableA2.tex


The recomputed ionic abundance ratios for M83 (NGC 5236) are given in Table A2.
{As in R07, values of $T_e = 8000$~K and $N_e = 100$~cm$^{-3}$ were used.}
This high-metallicity, nearly face-on spiral galaxy has relatively low excitation 
in its inner galaxy H~{\sc ii} regions, as seen in the excitation indicators 
Ne$^{++}$/Ne$^+$ and S$^{3+}$/S$^{++}$. 
Both of these show positive gradients 
with $R_{\rm G}$, with the former statistically significant: 

\vskip0.1truein

\noindent
Ne$^{++}$/Ne$^{+}$ = 0.049$\pm$0.015~+~(0.017$\pm$0.005)~$R_{\rm G}$,

\vskip0.1truein

\noindent
where $R_{\rm G}$ is in kpc.
This is most likely caused by metallicity decreasing with $R_{\rm G}$, 
where the lower metallicity causes a harder ionizing spectrum in the exciting stars.
Compared with the results in R07, the Ne$^+$/S$^{++}$ ratio is substantially lower,
resulting in the Ne/S ratio also being substantially lower since these 
are the dominant ions in H~{\sc ii} regions of Solar and higher than Solar metallicity.
Even so, all M83 Ne/S ratios are higher than the benchmark M42 ratio.  
The (Ne$^+$ + Ne$^{++}$)/(S$^{++}$ + S$^{++}$) gradient with $R_{\rm G}$ is similar
to that determined in R07.
The likely reason for the existance of a (Ne$^+$ + Ne$^{++}$)/(S$^{++}$ + S$^{++}$) gradient 
is that the high-metallicity H~{\sc ii} regions contain substantial amounts of uncounted S$^+$,
although it is also possible that the interstellar dust contains the missing sulphur (R07).

\subsection{M33}

The recomputed ionic abundance ratios for M33 (NGC 598) are given in Table A3.

In our study of M33 (R08), we measured  significant  gradients in the degree of ionization~--
Ne$^{++}$/Ne$^+$ and S$^{3+}$/S$^{++}$~ -- versus {$R_{\rm G}$}. 
The basic conclusions still hold qualitatively here.
For the entries in Table~A3, as previously, we adopt a value for all the M33
H~{\sc ii} regions of $T_e$~= 8000~K and $N_e$~= 100~cm$^{-3}$ (R08, section 3).
We present the variation of Ne$^{++}$/Ne$^+$ with $R_{\rm G}$
in Fig.~A1 using the values from our updated table of derived
parameters of the H~{\sc ii} regions in M33 (Table A3). The error
values here, as well as for the values in Tables A1 and A2,
represent the propagated uncertainties in the flux measurement and 
do not include the systematic uncertainties.
A positive correlation of the excitation with $R_{\rm G}$ (in kpc) 
is given by the linear least-squares fit:

\vskip0.1truein

\noindent
Ne$^{++}$/Ne$^{+}$ = $-$0.66$\pm$0.39~+~(0.623$\pm$0.117)~$R_{\rm G}$.

\vskip0.1truein

\noindent
For all the linear least-squares fits in this paper,
each point is given equal weight
because systematic uncertainties exceed the flux measurement
uncertainties, as discussed earlier.
The positive correlation of Ne$^{++}$/Ne$^{+}$
with $R_{\rm G}$ as measured by the slope is statistically 
significant by the criterion that it
exceeds the 3~$\sigma$ uncertainty.
Similar results are shown in Fig.~A2 for the S$^{3+}$/S$^{++}$ ratio.

In R08, we had made measurements of the H(7-6) flux for 16
of the H~{\sc ii} regions in M33. We used those measurements
to determine the heavy element abundances Ne/H and S/H. 
Our updated linear least-squares fit of  log~(Ne/H) versus $R_{\rm G}$
is given by

\vskip0.1truein

\noindent
log~(Ne/H)~= $-$4.04$\pm$0.05~$-$~(0.041$\pm$0.016)~$R_{\rm G}$

\noindent
and is shown in Fig.~A3.

\noindent
Our updated linear least-squares fit to log~(S/H) versus $R_{\rm G}$ is given by 
\vskip0.1truein

\noindent
log~(S/H)~= $-$5.11$\pm$0.06~$-$~(0.056$\pm$0.020)~$R_{\rm G}$.
\vskip0.1truein

Compared to the log~(Ne/H) fit in R08, the updated least-squares fit for log~(Ne/H) 
has a significantly smaller slope.
The reason for the change is that there is a substantial ionization gradient in M33, 
with the abundance of Ne$^{++}$/Ne$^+$ increasing with $R_{\rm G}$.
With the updated atomic data, the abundance ratio Ne$^+$/H$^+$ decreased by $\sim 10$ per cent 
but the abundance ratio Ne$^{++}$/H$^+$ {\it increased} by $\sim 37$ per cent.
(The abundance ratio S$^{++}$/H$^+$ also increased by $\sim 33$ per cent, 
which is why the Ne/S ratios changed so little for the highly excited H~{\sc ii} regions.)
It is clear that there are additional uncertainties in any gradient computation 
that are due to uncertainties in the atomic data; 
it is beyond the scope of this paper to discuss such uncertainties further.

Ionization increasing with $R_{\rm G}$ is usually attributed to metallicity decreasing
with $R_{\rm G}$, because stars with lower metallicity have harder ionizing spectra 
(Section 4, Fig. 3).
The reader is referred to R08 for further discussion.


\begin{figure}
\resizebox{8.4cm}{!}{\includegraphics{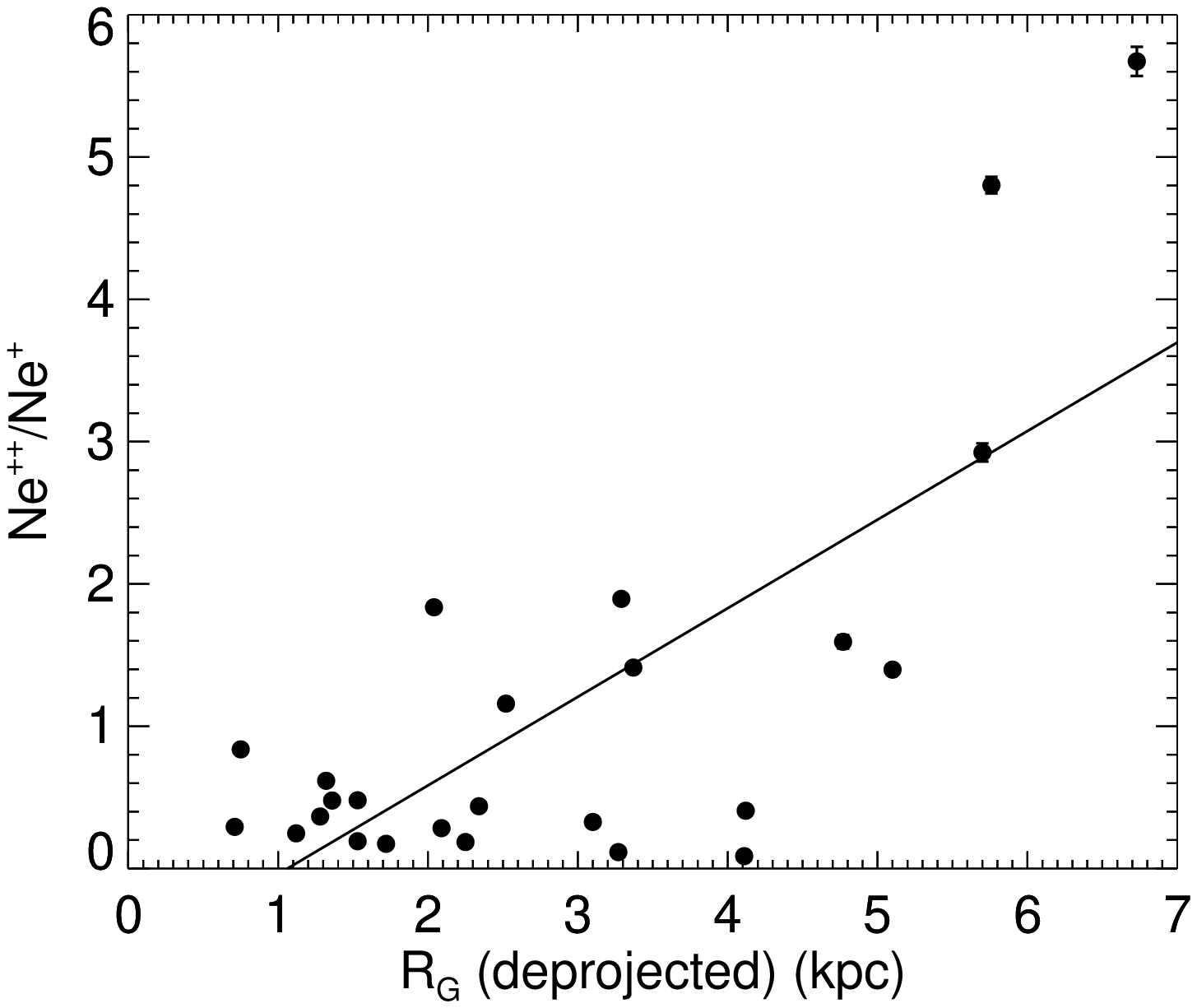} }
\vskip0.1truein
\caption[]{The ionic abundance ratio
Ne$^{++}$/Ne$^+$  versus $R_{\rm G}$ in M33.
This is derived from the measured line flux ratios for 
the 25 H~{\sc ii} regions in Table~A3.
The solid line is the linear least-squares fit
Ne$^{++}$/Ne$^{+}$ = $-$0.66$\pm$0.39~+~(0.623$\pm$0.117)~$R_{\rm G}$.
The error bars, which are smaller than the symbol size for most positions, 
are for the propagated measurement uncertainties 
and do not include the systematic uncertainties (see text).}
\end{figure}


\begin{figure}
\resizebox{8.4cm}{!}{\includegraphics{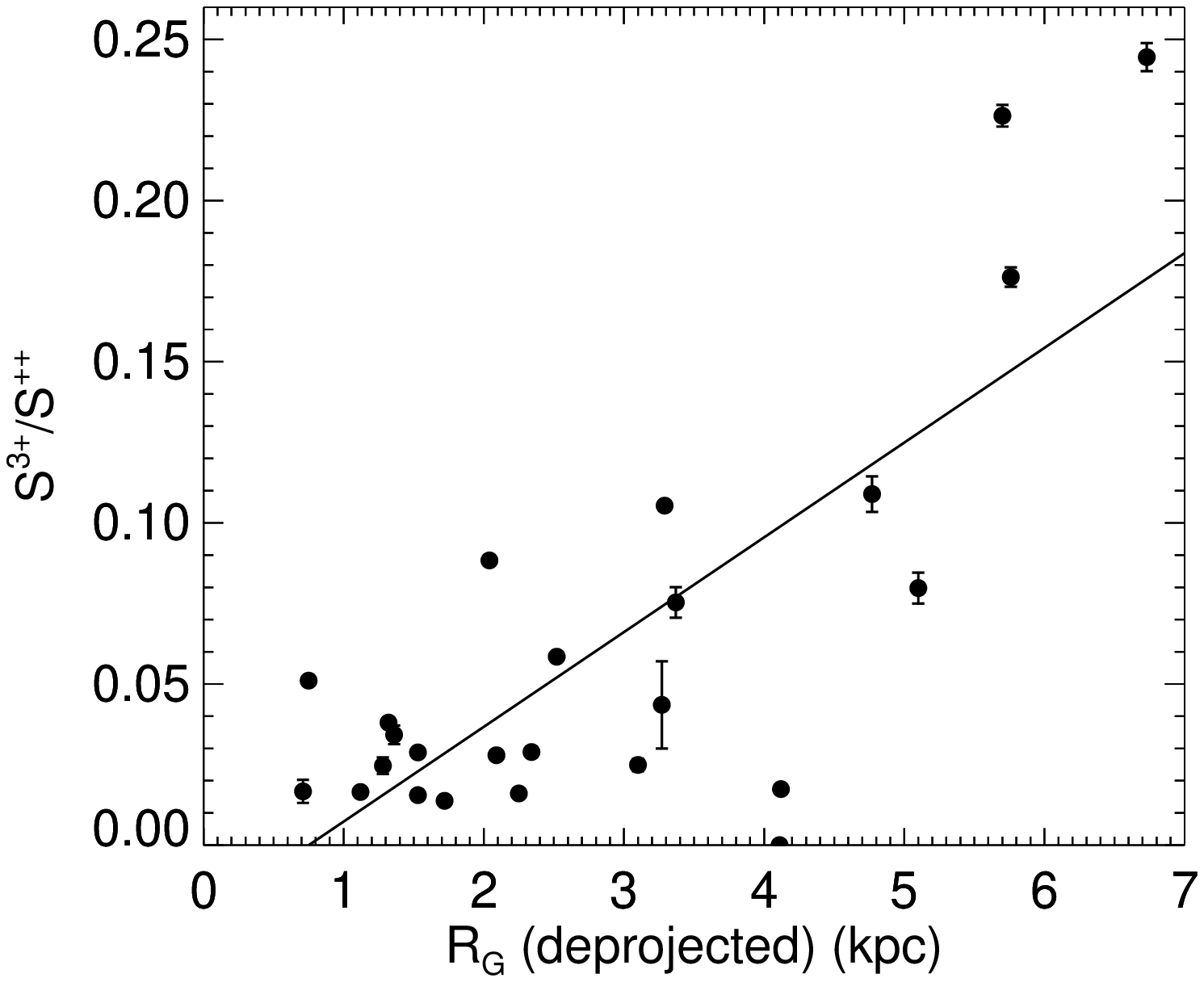} }
\vskip0.1truein
\caption[]{The ionic abundance ratio S$^{3+}$/S$^{++}$ vs.\ $R_{\rm G}$ in M33.
This is derived from the measured line flux ratios for 
the same 25 H~{\sc ii} regions as in Fig. A1. 
The solid line is the linear least-squares fit
S$^{3+}$/S$^{++}$ = $-$0.022$\pm$0.017~+~(0.029$\pm$0.005)~$R_{\rm G}$.
Error bars here are for the propagated measurement
uncertainties and do not include the systematic uncertainties (see text).}

\end{figure}


\begin{figure}
\resizebox{8.4cm}{!}{\includegraphics{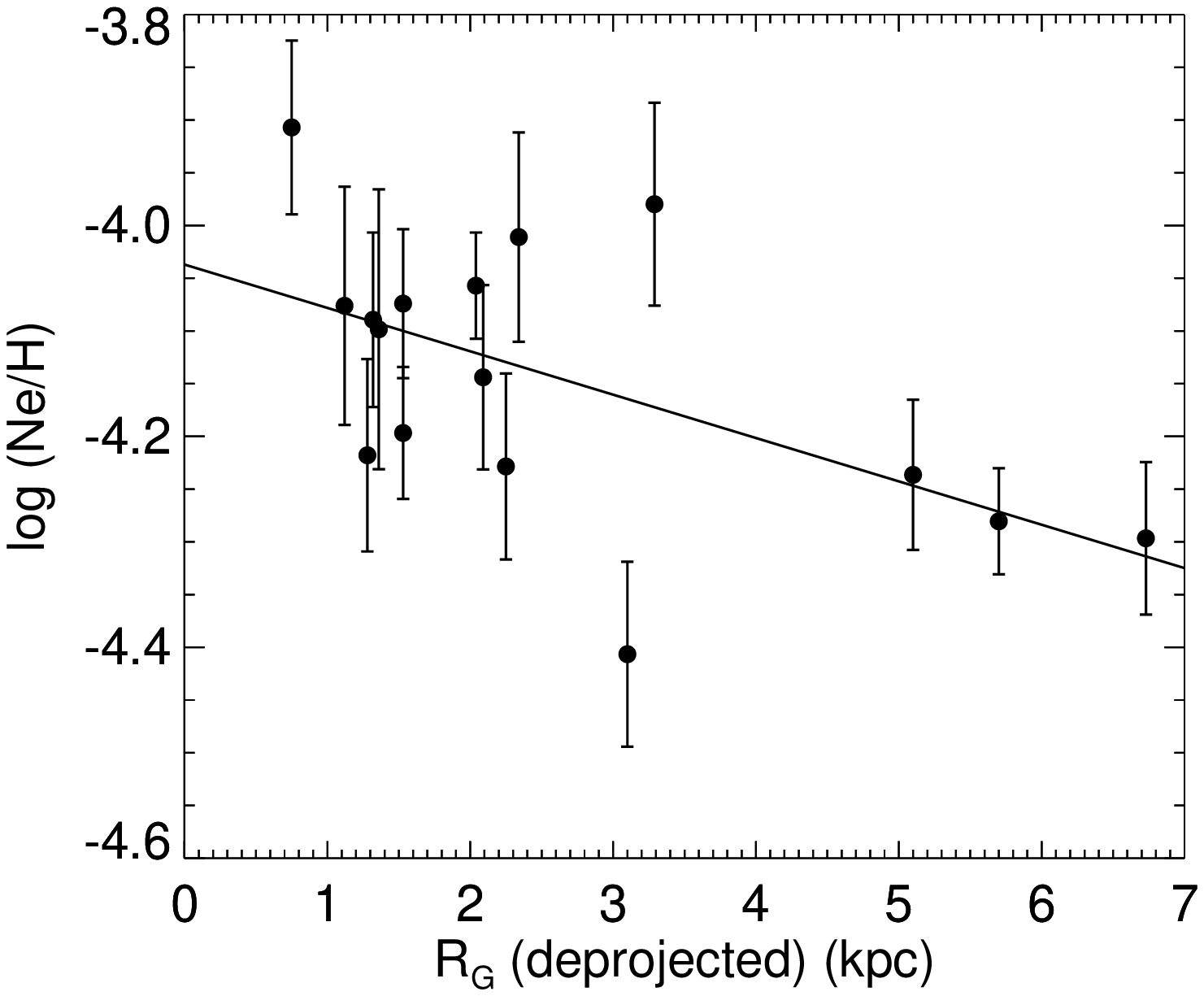} }
\vskip0.1truein
\caption[]{Plot of our updated computation of log~(Ne/H) versus $R_{\rm G}$ 
for the 16 sources in M33
with statistically significant H(7-6) measurements (Table A3).
The solid line is the  linear least-squares fit 
log~(Ne/H)~= $-4.04\pm0.05 -(0.041\pm0.016)$~$R_{\rm G}$.
}
\end{figure}

\input tableA3.tex


\subsection{Blue compact dwarf galaxies}

\input tableA4.tex

The recomputed ionic abundance ratios for the blue, compact dwarf galaxies 
from W08 are given in Table A4.
We employ the same semi-empirical analysis code for 
the line fluxes in their table 3 to derive the various ionic abundance ratios.
Only the galaxies where all four of the lines [S~{\sc iv}], [Ne~{\sc ii}], [Ne~{\sc iii}], and [S~{\sc iii}] 
were measured with good signal/noise were considered. 
Values of $T_e$ equal to the optically derived values found in Table 2 of W08 
and $N_e = 100$~cm$^{-3}$ were used; however, we note that the ionic ratios in this table 
are very insensitive to the choice of $N_e$ and $T_e$.

\subsection{Nearby massive H~{II} regions}

\input tableA5.tex

The recomputed ionic abundance ratios for the massive H~{\sc ii} regions 
from Table 2 of L08 are given in Table A5.
L08 described their observations of numerous positions in the Galactic source NGC 3603, 
the extremely massive H~{\sc ii} region 30 Doradus in the Large Magellanic Cloud (LMC), 
and N66 (NGC 346), which is the largest H~{\sc ii} region 
in the Small Magellanic Cloud (SMC). 
In our Table A5, we take the positions with the best signal/noise ratios 
and list the Ne$^{++}$/Ne$^+$, S$^{3+}$/S$^{++}$, and Ne/S ratios derived using
the same $T_e$- and $N_e$-values as L08 had used 
(10,000~K and 1000~cm$^{-3}$ for NGC~3603; 
10,000~K and 100~cm$^{-3}$ for 30 Dor; 
and 12,500~K and 100 cm$^{-3}$ for N66). 
The median (average) Ne/S values we obtain for NGC~3603, 30 Dor, and N66
are 10.6 (13.1~$\pm$2.6), 11.2 (11.2~$\pm$0.2), and 10.3 (10.4~$\pm$0.2) respectively.

\end{appendix}

\bsp

\label{lastpage}

\end{document}

%% file: table1.tex

\setcounter{table}{0}
\begin{table*}
\centering
\begin{minipage}{145mm}
\caption{H~{\sc ii} regions observed in NGC 6822.}
\vskip-0.1truein
\begin{tabular}{@{}lclcccc@{}}
\hline
H~{\sc ii} Region  & $R_{\rm G}$ & ~~RA~~~~(J2000)  & DEC & \#Parallel/Step & \#Perp./Step & Aperture \\
                   & (kpc)       &                &     &                 &             & (arcsec$^2$) \\ 
\hline
KD 2e  & 1.01  & 19 44 31.0 & -14 48 24 & 2 / 3.45$''$ & 5 / 2.3$''$ & 245 \\

Hu I  & 1.37  & 19 44 31.6 & -14 42 00 & 3 / 5.9$''$ & 5 / 2.65$''$ & 368 \\

KD 5e  & 0.874  & 19 44 32.9 & -14 47 29 & 2 / 3.45$''$ & 2 / 2.3$''$ & 123 \\

Hu III  & 1.28  & 19 44 34.0 & -14 42 21 & 3 / 5.9$''$ & 5 / 2.65$''$ & 368 \\

KD 10  & 0.783  & 19 44 48.3 & -14 44 18 & 2 / 3.45$''$ & 5 / 2.3$''$ & 245 \\

KD 18  & 0.909  & 19 44 52.1 & -14 51 56 & 2 / 3.45$''$ & 2 / 2.3$''$ & 107 \\

Hu V  & 1.03  & 19 44 52.8 & -14 43 12 & 3 / 5.9$''$ & 5 / 2.65$''$ & 429 \\

KD 28e  & 0.065  & 19 44 57.1 & -14 47 49 & 2 / 3.45$''$ & 2 / 2.3$''$ & 123 \\

Hu X  & 1.20  & 19 45 05.2 & -14 43 12 & 3 / 5.9$''$ & 5 / 2.65$''$ & 337 \\
\hline
\end{tabular}
\end{minipage}
\end{table*}

%% file: table2.tex

\setcounter{table}{1}
\begin{table*}
\centering
\begin{minipage}{165mm}
\caption{NGC 6822 line measurements. Intensities less than 3$\sigma$ were not used in the analysis.
The uncertainties in the full width at half maximum (FWHM) and radial velocity of the brightest lines are $\sim 10 - 20$ km s$^{-1}$ 
(these uncertainties for {\it Spitzer} IRS line measurements are estimated from the dispersion in the FWHM of the brightest lines in R11 and the velocity measurements described by Simpson et al. 2007).
}
\begin{tabular}{@{}lccccr@{}}
\hline
Source & Line & Intensity & 1$\sigma$ error & FWHM & $V_{\rm helio}$ \\
& ($\mu$m) & \multicolumn{2}{c}{{(ergs cm$^{-2}$ s$^{-1}$ arcsec$^{-2}$)}} & (km s$^{-1}$) & (km s$^{-1}$) \\
\hline
 KD 2e& 10.5 & 1.87E-16 & 2.06E-17 & 601 & 14 \\
      & 12.4 & - & - & - & - \\
      & 12.8 & 2.12E-17 & 5.68E-18 & 363 & -118 \\
      & 15.6 & 2.05E-16 & 5.78E-18 & 379 & -67 \\
      & 18.7 & 1.18E-16 & 6.40E-18 & 406 & -90 \\

 Hu I & 10.5 & 1.67E-16 & 2.56E-17 & 405 & -96 \\
      & 12.4 & - & - & - & - \\
      & 12.8 & 1.12E-16 & 1.31E-17 & 376 & -42 \\
      & 15.6 & 3.91E-16 & 2.34E-17 & 534 & -100 \\
      & 18.7 & 2.81E-16 & 2.66E-17 & 499 & -84 \\

 KD 5e& 10.5 & 1.52E-16 & 1.31E-17 & 511 & -17 \\
      & 12.4 & - & - & - & - \\
      & 12.8 & 6.18E-17 & 5.96E-18 & 411 & -72 \\
      & 15.6 & 2.97E-16 & 4.59E-18 & 388 & -41 \\
      & 18.7 & 1.74E-16 & 9.02E-18 & 416 & -55 \\
 
 Hu III& 10.5 & 4.17E-17 & 1.17E-17 & 314 & -227 \\
      & 12.4 & - & - & - & - \\
      & 12.8 & 9.62E-17 & 1.64E-17 & 502 & -65 \\
      & 15.6 & 1.86E-16 & 1.45E-17 & 611 & -142 \\
      & 18.7 & 1.22E-16 & 1.41E-17 & 480 & -61 \\

 KD 10& 10.5 & 4.48E-17 & 8.78E-18 & 517 & 91 \\
      & 12.4 & - & - & - & - \\
      & 12.8 & 4.48E-17 & 4.72E-18 & 433 & -20 \\
      & 15.6 & 7.37E-17 & 8.45E-18 & 436 & -34 \\
      & 18.7 & 8.94E-17 & 3.44E-18 & 405 & -67 \\

KD 18 & 10.5 & 1.06E-16 & 7.19E-18 & 509 & -18 \\
      & 12.4 & - & - & - & - \\
      & 12.8 & 6.93E-17 & 5.01E-18 & 406 & -43 \\
      & 15.6 & 1.80E-16 & 4.20E-18 & 419 & -38 \\
      & 18.7 & 1.51E-16 & 4.80E-18 & 438 & -48 \\

 Hu V & 10.5 & 2.63E-15 & 4.00E-17 & 465 & -36 \\
      & 12.4 & 1.45E-16 & 1.25E-17 & 423 & -15 \\
      & 12.8 & 5.61E-16 & 1.28E-17 & 402 & -42 \\
      & 15.6 & 3.06E-15 & 4.32E-17 & 425 & -54 \\
      & 18.7 & 2.09E-15 & 5.25E-17 & 455 & -64 \\
 
 KD 28e& 10.5 & 9.19E-17 & 1.28E-17 & 480 & 14 \\
      & 12.4 & - & - & - & - \\
      & 12.8 & 5.04E-17 & 7.01E-18 & 342 & -67 \\
      & 15.6 & 1.35E-16 & 4.36E-18 & 377 & -50 \\
      & 18.7 & 1.02E-16 & 5.71E-18 & 426 & -61 \\

 Hu X & 10.5 & 6.67E-16 & 2.37E-17 & 479 & -50 \\
      & 12.4 & 7.55E-17 & 1.51E-17 & 436 & 48 \\
      & 12.8 & 3.21E-16 & 1.38E-17 & 385 & -30 \\
      & 15.6 & 1.16E-15 & 3.74E-17 & 453 & -45 \\
      & 18.7 & 9.80E-16 & 3.12E-17 & 455 & -41 \\

\hline
\end{tabular}
\end{minipage}
\end{table*}

%% file: table3.tex






\setcounter{table}{2}

\begin{landscape}
\begin{table}
\vskip2.0truein
\caption{~Derived Parameters for the H~{\sc ii} Regions in NGC 6822.}
\begin{tabular}{@{}lccccccccccccc@{}}
\hline
 Source & $R_{\rm G}$ & \underbar{Ne$^+$} & \underbar{Ne$^{++}$} & \underbar{S$^{++}$} & \underbar{S$^{3+}$} & \underbar{Ne$^+$} & \underbar{Ne$^{++}$} & \underbar{Ne$^{++}$} & \underbar{S$^{3+}$} & \underbar{Ne} & \underbar{$<$S$^{++}$$>$} & \underbar{$<$Ne$^{++}$$>$} & \underbar{$<$Ne$^{++}$$>$} \\
  &  & H$^+$ & H$^+$ & H$^+$ & H$^+$ & S$^{++}$ & S$^{++}$ & Ne$^+$ & S$^{++}$ & S & $<$Ne$^+$$>$ & $<$S$^{++}$$>$ & $<$S$^{3+}$$>$ \\

  & (kpc) & $($$\times${10$^{-6}$}$)$ & $($$\times${10$^{-6}$}$)$ & $($$\times${10$^{-6}$}$)$ & $($$\times${10$^{-8}$}$)$ &  &  &  &  &  &  &  & \\
\hline

  KD 2e &  1.01 &   -- &  -- & -- &  -- &   1.88 $\pm$ 0.51 & 11.82 $\pm$ 0.72 &  6.30 $\pm$ 1.70 &  0.258 $\pm$ 0.032 &  10.9 $\pm$ 0.7 &  4.84 $\pm$ 1.32 & 1.302 $\pm$ 0.079 &  5.04 $\pm$ 0.57 \\

   Hu I &  1.37 &   -- &  -- & -- & -- &   4.17 $\pm$ 0.62 &  9.48 $\pm$ 1.05 &  2.27 $\pm$ 0.30 &  0.097 $\pm$ 0.017 &  12.4 $\pm$ 1.3 &  2.18 $\pm$ 0.33 & 1.044 $\pm$ 0.116 & 10.77 $\pm$ 1.77 \\

  KD 5e &  0.874 &   -- &  -- & -- &  -- & 3.72 $\pm$ 0.41 & 11.68 $\pm$ 0.63 &  3.14 $\pm$ 0.31 &  0.143 $\pm$ 0.014 &  13.5 $\pm$ 0.7 &  2.44 $\pm$ 0.27 & 1.286 $\pm$ 0.069 &  9.01 $\pm$ 0.79 \\

 Hu III &  1.28 &   -- &  -- & -- &  -- &  8.26 $\pm$ 1.70 & 10.43 $\pm$ 1.44 &  1.26 $\pm$ 0.24 &  0.056 $\pm$ 0.017 &  17.7 $\pm$ 2.5$^a$ &  1.10 $\pm$ 0.23 & 1.148 $\pm$ 0.159 & 20.61 $\pm$ 6.02 \\

  KD 10 &  0.783 &  -- &  -- & -- &  -- & 5.25 $\pm$ 0.59 &  5.62 $\pm$ 0.68 &  1.07 $\pm$ 0.17 &  0.082 $\pm$ 0.016 &  10.1 $\pm$ 0.9 &  1.73 $\pm$ 0.19 & 0.619 $\pm$ 0.075 &  7.59 $\pm$ 1.72 \\

  KD 18 &  0.909 &   -- &  -- & -- &  -- & 4.80 $\pm$ 0.38 &  8.13 $\pm$ 0.32 &  1.69 $\pm$ 0.13 &  0.114 $\pm$ 0.009 &  11.6 $\pm$ 0.5 &  1.89 $\pm$ 0.15 & 0.895 $\pm$ 0.035 &  7.88 $\pm$ 0.57 \\

   Hu V &  1.03 &   5.06 $\pm$ 0.55 & 18.0 $\pm$ 1.9 & 1.80 $\pm$ 0.20 & 36.9 $\pm$ 4.0 &  2.81 $\pm$ 0.09 & 10.00 $\pm$ 0.29 &  3.56 $\pm$ 0.10 &  0.205 $\pm$ 0.006 &  10.6 $\pm$ 0.3 &  3.23 $\pm$ 0.11 & 1.102 $\pm$ 0.032 &  5.38 $\pm$ 0.11 \\

 KD 28e &  0.065 &   -- &  -- & -- &  -- &  5.15 $\pm$ 0.77 &  8.99 $\pm$ 0.57 &  1.75 $\pm$ 0.25 &  0.146 $\pm$ 0.022 &  12.3 $\pm$ 0.9 &  1.76 $\pm$ 0.26 & 0.990 $\pm$ 0.063 &  6.78 $\pm$ 0.96 \\

   Hu X &  1.20 &   5.55 $\pm$ 1.19 & 13.1 $\pm$ 2.8 & 1.62 $\pm$ 0.34 & 18.0 $\pm$ 3.8 &  3.42 $\pm$ 0.18 &  8.06 $\pm$ 0.36 &  2.36 $\pm$ 0.13 &  0.111 $\pm$ 0.005 &  10.3 $\pm$ 0.4 &  2.65 $\pm$ 0.14 & 0.888 $\pm$ 0.040 &  8.03 $\pm$ 0.38 \\

\hline
\end{tabular}
$^a$ Hu III was especially plagued by bad pixels in the vicinity of the sulphur lines at 10.5 and 18.7 \micron; this is probably the source of the large error.
\end{table}
\end{landscape}


%% file: table4.tex

\setcounter{table}{3}
\begin{table*}
\centering
\begin{minipage}{170mm}
\caption{Average and median Ne/S values versus metallicity.}
\vskip-0.1truein
\begin{tabular}{@{}lccccccccc@{}}
\hline 
       & \multicolumn{3}{c}{Average} & \ \  & \multicolumn{3}{c}{Median} && \\
\cline{2-4}  \cline{6-8} 
Object & Ne/H & S/H$^a$ & Ne/S$^b$ && Ne/H & S/H$^a$ & Ne/S$^b$ & Z/Z$_{\odot}$ & 12+log(O/H) \\
  & $($$\times${10$^{-6}$}$)$ & $($$\times${10$^{-6}$}$)$ && & $($$\times${10$^{-6}$}$)$ & $($$\times${10$^{-6}$}$)$ & & \\
\hline
NGC 6822 & 20.9$\pm$2.2 & 1.99$\pm$0.19 & 12.2$\pm$0.8 && 20.9 & 1.99 & 11.6 & 0.3$^c$ & 8.06$\pm$0.04$^d$ \\
N66 (SMC)  & 24.9$\pm$4.4 & 2.39$\pm$0.42 & 10.4$\pm$0.2 && 25.7 & 2.54 & 10.3 & 0.2$\pm$0.1$^e$ & 8.0 \\
30 Dor (LMC) & 71.1$\pm$2.2 & 6.36$\pm$0.27 & 11.2$\pm$0.2 && 70.1 & 6.22 & 11.2 & 0.6$^f$ & 8.33$\pm$0.02$^g$ \\
M33 & 74.9$\pm$5.6 & 5.86$\pm$0.53 & 13.7$\pm$0.6 && 75.8 & 5.69 & 12.8 & -  & 8.53$\pm$0.05 - (0.054$\pm$0.011)$R_{\rm G}$$^h$ \\
NGC 3603 & 99.3$\pm$4.0 & 8.04$\pm$1.25 & 13.1$\pm$2.6 && 102.3 & 8.88 & 10.6 & 0.5 - 0.7  & 8.39 - 8.52$^i$ \\
M42 & 101.9$\pm$2.9 & 9.61$\pm$0.31 & 10.5$\pm$0.2 && 101.8 & 9.70 & 10.4 & 0.81 & 8.60$^j$ \\
\hline
\end{tabular}
\\
$^a$S/H$ = ($S$^{++}$ + S$^{3+}$)/H$^+$.  \\
$^b$Ne/S$ = ($Ne$^+$ + Ne$^{++}$)/(S$^{++}$ + S$^{3+}$).  \\
$^c$Galametz et al. 2010. \\
$^d$Hern\'{a}ndez-Mart\'{i}nez 2009. \\
$^e$Bouret et al. 2003.  \\
$^f$Lebouteiller et al. 2008. \\
$^g$Peimbert 2003. \\
$^h$Magrini et al. 2007. \\
$^i$Melnick et al. 1989; Tapia et al. 2001; Garc\'ia-Rojas et al. 2006. \\
$^j$Rubin et al. 1991. \\
\end{minipage}
\end{table*}

%% file: tableA1.tex



\setcounter{table}{0}
\begin{table*}
\centering
\begin{minipage}{250mm}
\caption{Derived parameters for the positions in the M42 H~{\sc ii} region.}

\begin{tabular}{@{}lcccccccccc@{}}
\hline
Chex & $N_e$[S~{\sc iii}] & \underbar{Ne$^+$} & \underbar{Ne$^{++}$} & \underbar{S$^{++}$} & \underbar{S$^{3+}$}  & \underbar{Ne$^{++}$} & \underbar{S$^{3+}$} & \underbar{Ne$^a$}  \\
 &  & H$^+$ & H$^+$ & H$^+$ & H$^+$  & Ne$^+$ & S$^{++}$ & S  \\
 & (cm$^{-3}$) & $($$\times${10$^{-6}$}$)$ & $($$\times${10$^{-6}$}$)$ & $($$\times${10$^{-6}$}$)$ & $($$\times${10$^{-8}$}$)$ & $($$\times${10$^{-3}$}$)$ & $($$\times${10$^{-3}$}$)$ &  \\
\hline

    I4 & 1433 & 101.0 $\pm$  8.9 & 11.750 $\pm$  1.038 & 9.82 $\pm$ 0.87 & 13.37 $\pm$ 1.21 & 116.37 $\pm$ 1.02 & 13.62 $\pm$ 0.28 & 11.3 $\pm$ 0.1 \\ 
    I3 &  990 &  99.2 $\pm$  8.8 &  9.175 $\pm$  0.818 & 9.67 $\pm$ 0.86 &  8.38 $\pm$ 0.76 &  92.47 $\pm$ 0.74 &  8.66 $\pm$ 0.18 & 11.1 $\pm$ 0.1 \\ 
    I2 &  901 &  99.5 $\pm$  8.6 &  5.555 $\pm$  0.484 & 9.67 $\pm$ 0.84 &  4.79 $\pm$ 0.43 &  55.81 $\pm$ 0.54 &  4.95 $\pm$ 0.13 & 10.8 $\pm$ 0.1 \\ 
    I1 &  705 &  94.6 $\pm$  8.7 &  4.378 $\pm$  0.398 & 9.51 $\pm$ 0.86 &  3.42 $\pm$ 0.34 &  46.26 $\pm$ 0.97 &  3.60 $\pm$ 0.16 & 10.4 $\pm$ 0.2 \\ 
    M1 &  574 &  95.9 $\pm$  8.3 &  5.549 $\pm$  0.482 & 9.64 $\pm$ 0.83 &  3.67 $\pm$ 0.34 &  57.88 $\pm$ 0.55 &  3.81 $\pm$ 0.13 & 10.5 $\pm$ 0.1 \\ 
    M2 &  444 &  89.0 $\pm$  8.6 &  4.055 $\pm$  0.386 & 9.06 $\pm$ 0.86 &  2.51 $\pm$ 0.26 &  45.54 $\pm$ 0.94 &  2.77 $\pm$ 0.12 & 10.2 $\pm$ 0.2 \\ 
    M3 &  406 &  85.4 $\pm$  9.0 &  2.845 $\pm$  0.299 & 9.08 $\pm$ 0.95 &  1.62 $\pm$ 0.20 &  33.30 $\pm$ 0.60 &  1.78 $\pm$ 0.13 &  9.7 $\pm$ 0.1 \\ 
    M4 &  386 & 100.1 $\pm$  9.0 &  2.090 $\pm$  0.191 & 9.97 $\pm$ 0.90 &  1.64 $\pm$ 0.27 &  20.88 $\pm$ 0.36 &  1.64 $\pm$ 0.22 & 10.2 $\pm$ 0.1 \\ 
  V1-1 &  249 & 101.5 $\pm$  9.3 &  1.843 $\pm$  0.176 & 8.57 $\pm$ 0.79 &  0.94 $\pm$ 0.10 &  18.17 $\pm$ 0.48 &  1.10 $\pm$ 0.06 & 12.0 $\pm$ 0.1 \\ 
  V1-2 &  244 & 105.1 $\pm$  9.4 &  1.979 $\pm$  0.181 & 8.79 $\pm$ 0.79 &  1.48 $\pm$ 0.22 &  18.83 $\pm$ 0.34 &  1.68 $\pm$ 0.20 & 12.2 $\pm$ 0.1 \\ 
  V1-3 &  207 & 109.6 $\pm$ 10.8 &  1.921 $\pm$  0.191 & 9.30 $\pm$ 0.91 &  1.84 $\pm$ 0.48 &  17.53 $\pm$ 0.36 &  1.98 $\pm$ 0.47 & 12.0 $\pm$ 0.2 \\ 
  V2-1 &  342 & 102.7 $\pm$  9.8 &  0.826 $\pm$  0.084 & 5.62 $\pm$ 0.54 &  1.15 $\pm$ 0.13 &   8.05 $\pm$ 0.30 &  2.05 $\pm$ 0.12 & 18.4 $\pm$ 0.1 \\ 
  V2-2 &  340 & 109.8 $\pm$  9.8 &  0.806 $\pm$  0.073 & 6.05 $\pm$ 0.55 &  1.19 $\pm$ 0.14 &   7.34 $\pm$ 0.12 &  1.97 $\pm$ 0.16 & 18.3 $\pm$ 0.2 \\ 
  V2-3 &  280 &  98.4 $\pm$  8.7 &  0.710 $\pm$  0.068 & 5.42 $\pm$ 0.48 &  1.00 $\pm$ 0.17 &   7.22 $\pm$ 0.25 &  1.84 $\pm$ 0.26 & 18.3 $\pm$ 0.1 \\ 
  V3-1 &   65 &  54.2 $\pm$  4.9 &  2.071 $\pm$  0.212 & 2.64 $\pm$ 0.23 &  2.94 $\pm$ 0.35 &  38.23 $\pm$ 2.43 & 11.16 $\pm$ 0.94 & 21.1 $\pm$ 0.7 \\ 
  V3-2 &   $< 100$ &  58.4 $\pm$  5.3 &  2.147 $\pm$  0.219 & 2.74 $\pm$ 0.24 &  3.20 $\pm$ 0.64 &  36.75 $\pm$ 2.32 & 11.99 $\pm$ 2.09 & 21.8 $\pm$ 0.7 \\ 
  V3-3 &  $< 100$ &  53.3 $\pm$  4.9 &  1.834 $\pm$  0.191 & 2.49 $\pm$ 0.22 &  2.83 $\pm$ 0.24 &  34.44 $\pm$ 2.32 & 11.36 $\pm$ 0.16 & 21.9 $\pm$ 0.7 \\ 

\hline
\end{tabular}
\end{minipage}
\noindent $^a$ (Ne$^+$ + Ne$^{++}$)/(S$^{++}$ + S$^{3+}$). See text for correction for S$^+$.
\end{table*}


%% file: tableA2.tex



\setcounter{table}{1}
\begin{landscape}
\begin{table}
\caption{Derived parameters for the H~{\sc ii} regions in M83.}

\begin{tabular}{lccccccccc}
\hline
Source & $R_{\rm G}$ &  \underbar{Ne$^+$} & \underbar{Ne$^{++}$} & \underbar{Ne$^{++}$} & \underbar{S$^{3+}$} & \underbar{Ne} & \underbar{$<$S$^{++}$$>$} & \underbar{$<$Ne$^{++}$$>$} & \underbar{$<$Ne$^{++}$$>$} \\
 & (kpc)  & S$^{++}$ & S$^{++}$ & Ne$^+$ & S$^{++}$ & S & $<$Ne$^+$$>$ & $<$S$^{++}$$>$ & $<$S$^{3+}$$>$ \\
\hline
RK275 &  5.16 &  15.3 $\pm$ 0.2 &  1.47 $\pm$ 0.05 & 0.0961 $\pm$ 0.0034 & 0.0103 $\pm$ 0.0013 &  16.6 $\pm$ 0.2 & 0.593 $\pm$ 0.008 & 0.162 $\pm$ 0.006 & 15.69 $\pm$ 2.00 \\
 RK268 &  4.51 &  16.6 $\pm$ 0.9 &  3.07 $\pm$ 0.17 & 0.1850 $\pm$ 0.0099 & 0.0728 $\pm$ 0.0140 &  18.3 $\pm$ 0.9 & 0.548 $\pm$ 0.031 & 0.338 $\pm$ 0.018 &  4.64 $\pm$ 0.89 \\
 RK266 &  4.29 &  15.0 $\pm$ 0.4 &  1.96 $\pm$ 0.23 & 0.1311 $\pm$ 0.0149 & 0.0144 $\pm$ 0.0037 &  16.7 $\pm$ 0.4 & 0.606 $\pm$ 0.015 & 0.216 $\pm$ 0.025 & 15.04 $\pm$ 4.25 \\
 RK230 &  2.50 &  16.8 $\pm$ 0.6 &  2.86 $\pm$ 0.16 & 0.1701 $\pm$ 0.0078 & 0.0130 $\pm$ 0.0020 &  19.4 $\pm$ 0.7 & 0.540 $\pm$ 0.020 & 0.315 $\pm$ 0.017 & 24.29 $\pm$ 3.81 \\
 deV10 &  2.34 &  18.8 $\pm$ 0.6 &  1.54 $\pm$ 0.18 & 0.0819 $\pm$ 0.0093 & 0.0076 $\pm$ 0.0005 &  20.1 $\pm$ 0.7 & 0.484 $\pm$ 0.016 & 0.169 $\pm$ 0.020 & 22.31 $\pm$ 2.83 \\
 RK213 &  2.07 &  18.2 $\pm$ 0.2 &  0.80 $\pm$ 0.04 & 0.0442 $\pm$ 0.0021 & 0.0025 $\pm$ 0.0004 &  19.0 $\pm$ 0.2 & 0.498 $\pm$ 0.004 & 0.089 $\pm$ 0.004 & 35.89 $\pm$ 6.21 \\
 deV13 &  2.15 &  18.4 $\pm$ 0.2 &  1.31 $\pm$ 0.11 & 0.0710 $\pm$ 0.0058 & 0.0059 $\pm$ 0.0008 &  19.6 $\pm$ 0.2 & 0.492 $\pm$ 0.006 & 0.144 $\pm$ 0.012 & 24.44 $\pm$ 3.80 \\
 RK211 &  2.23 &  16.6 $\pm$ 0.7 &  1.12 $\pm$ 0.07 & 0.0675 $\pm$ 0.0033 & 0.0059 $\pm$ 0.0006 &  17.6 $\pm$ 0.7 & 0.547 $\pm$ 0.022 & 0.123 $\pm$ 0.008 & 20.87 $\pm$ 2.27 \\
 RK209 &  2.02 &  16.9 $\pm$ 0.1 &  1.30 $\pm$ 0.03 & 0.0768 $\pm$ 0.0019 & 0.0045 $\pm$ 0.0005 &  18.1 $\pm$ 0.2 & 0.537 $\pm$ 0.005 & 0.143 $\pm$ 0.004 & 31.47 $\pm$ 3.61 \\
 RK201 &  4.00 &  17.2 $\pm$ 1.7 &  2.37 $\pm$ 0.32 & 0.1374 $\pm$ 0.0134 & 0.0271 $\pm$ 0.0048 &  19.1 $\pm$ 1.8 & 0.527 $\pm$ 0.051 & 0.261 $\pm$ 0.035 &  9.62 $\pm$ 1.72 \\
 RK198 &  2.18 &  15.1 $\pm$ 0.3 &  1.68 $\pm$ 0.05 & 0.1111 $\pm$ 0.0032 & 0.0112 $\pm$ 0.0013 &  16.6 $\pm$ 0.4 & 0.600 $\pm$ 0.013 & 0.185 $\pm$ 0.006 & 16.51 $\pm$ 1.95 \\
 deV22 &  1.91 &  20.5 $\pm$ 0.3 &  1.20 $\pm$ 0.03 & 0.0586 $\pm$ 0.0014 & 0.0048 $\pm$ 0.0008 &  21.6 $\pm$ 0.3 & 0.443 $\pm$ 0.006 & 0.132 $\pm$ 0.003 & 27.73 $\pm$ 4.45 \\
 RK172 &  0.95 &  21.0 $\pm$ 1.2 &  2.29 $\pm$ 0.38 & 0.1089 $\pm$ 0.0177 & 0.0156 $\pm$ 0.0041 &  22.9 $\pm$ 1.3 & 0.432 $\pm$ 0.024 & 0.252 $\pm$ 0.042 & 16.19 $\pm$ 4.91 \\
 RK154 &  4.41 &  15.0 $\pm$ 0.5 &  2.25 $\pm$ 0.13 & 0.1503 $\pm$ 0.0080 & 0.0143 $\pm$ 0.0021 &  17.0 $\pm$ 0.6 & 0.606 $\pm$ 0.020 & 0.248 $\pm$ 0.015 & 17.41 $\pm$ 2.70 \\
 deV28 &  0.77 &  21.3 $\pm$ 1.0 &  2.03 $\pm$ 0.25 & 0.0954 $\pm$ 0.0111 & 0.0124 $\pm$ 0.0036 &  23.0 $\pm$ 1.1 & 0.427 $\pm$ 0.020 & 0.224 $\pm$ 0.027 & 18.04 $\pm$ 5.61 \\
 deV31 &  0.46 &  26.7 $\pm$ 0.2 &  1.32 $\pm$ 0.06 & 0.0494 $\pm$ 0.0024 & 0.0038 $\pm$ 0.0003 &  27.9 $\pm$ 0.2 & 0.341 $\pm$ 0.003 & 0.145 $\pm$ 0.007 & 38.34 $\pm$ 3.47 \\
 RK137 &  0.57 &  18.6 $\pm$ 0.2 &  0.93 $\pm$ 0.04 & 0.0503 $\pm$ 0.0023 & 0.0026 $\pm$ 0.0006 &  19.5 $\pm$ 0.2 & 0.489 $\pm$ 0.005 & 0.103 $\pm$ 0.005 & 38.98 $\pm$ 9.33 \\
 RK135 &  4.05 &  16.4 $\pm$ 1.4 &  0.89 $\pm$ 0.13 & 0.0541 $\pm$ 0.0071 & 0.0083 $\pm$ 0.0014 &  17.2 $\pm$ 1.4 & 0.553 $\pm$ 0.046 & 0.098 $\pm$ 0.014 & 11.74 $\pm$ 2.26 \\
 RK120 &  2.48 &  17.2 $\pm$ 0.2 &  1.38 $\pm$ 0.04 & 0.0800 $\pm$ 0.0021 & 0.0094 $\pm$ 0.0004 &  18.4 $\pm$ 0.2 & 0.528 $\pm$ 0.006 & 0.152 $\pm$ 0.004 & 16.07 $\pm$ 0.81 \\
 RK110 &  1.38 &  22.8 $\pm$ 0.3 &  1.32 $\pm$ 0.05 & 0.0577 $\pm$ 0.0024 & 0.0045 $\pm$ 0.0010 &  24.0 $\pm$ 0.3 & 0.398 $\pm$ 0.005 & 0.145 $\pm$ 0.006 & 32.25 $\pm$ 6.94 \\
  RK86 &  2.93 &  17.9 $\pm$ 0.2 &  1.12 $\pm$ 0.02 & 0.0626 $\pm$ 0.0010 & 0.0056 $\pm$ 0.0004 &  18.9 $\pm$ 0.2 & 0.508 $\pm$ 0.004 & 0.123 $\pm$ 0.002 & 22.07 $\pm$ 1.65 \\
  RK69 &  1.89 &  21.2 $\pm$ 0.6 &  1.77 $\pm$ 0.10 & 0.0833 $\pm$ 0.0045 & 0.0084 $\pm$ 0.0019 &  22.8 $\pm$ 0.6 & 0.428 $\pm$ 0.011 & 0.195 $\pm$ 0.011 & 23.25 $\pm$ 5.37 \\
deV52+RK70 &  1.92 &  19.2 $\pm$ 0.3 &  1.26 $\pm$ 0.05 & 0.0657 $\pm$ 0.0027 & 0.0042 $\pm$ 0.0006 &  20.4 $\pm$ 0.3 & 0.472 $\pm$ 0.007 & 0.139 $\pm$ 0.006 & 33.55 $\pm$ 5.09 \\
  RK20 &  4.28 &  22.2 $\pm$ 1.4 &  3.20 $\pm$ 0.32 & 0.1439 $\pm$ 0.0124 & 0.0129 $\pm$ 0.0032 &  25.1 $\pm$ 1.5 & 0.408 $\pm$ 0.025 & 0.353 $\pm$ 0.035 & 27.35 $\pm$ 6.97 \\

\hline
\end{tabular}
\end{table}
\end{landscape}


%% file: tableA3.tex

\setcounter{table}{2}
\begin{landscape}
\begin{table}
\hskip-0.7truein
\begin{minipage}{215mm}
\vskip1.5truein
\caption{Derived parameters for the H~{\sc ii} regions in M33.}

\begin{tabular}{lccccccccccccc}
\hline
Source & $R_{\rm G}$ & \underbar{Ne$^+$} & \underbar{Ne$^{++}$} & \underbar{S$^{++}$} & \underbar{S$^{3+}$} & \underbar{Ne$^+$} & \underbar{Ne$^{++}$} & \underbar{Ne$^{++}$} & \underbar{S$^{3+}$} & \underbar{Ne} & \underbar{$<$S$^{++}$$>$} & \underbar{$<$Ne$^{++}$$>$} & \underbar{$<$Ne$^{++}$$>$} \\
 &  & H$^+$ & H$^+$ & H$^+$ & H$^+$ & S$^{++}$ & S$^{++}$ & Ne$^+$ & S$^{++}$ & S & $<$Ne$^+$$>$ & $<$S$^{++}$$>$ & $<$S$^{3+}$$>$ \\
  & (kpc) & $($$\times${10$^{-6}$}$)$ & $($$\times${10$^{-6}$}$)$ & $($$\times${10$^{-6}$}$)$ & $($$\times${10$^{-8}$}$)$ &  &  &  &  &  &  &  & \\
\hline

   280 &  5.76 &   -- &  -- &  -- & -- &   2.59 $\pm$ 0.04 & 12.46 $\pm$ 0.14 &  4.803 $\pm$ 0.058 & 0.1763 $\pm$ 0.0030 &  12.8 $\pm$ 0.1 & 3.500 $\pm$ 0.055 & 1.372 $\pm$ 0.015 &  7.78 $\pm$ 0.11 \\
   230 &  4.11 &    -- &  -- &  -- &   -- &  20.30 $\pm$ 1.40 &  1.77 $\pm$ 0.53 &  0.087 $\pm$ 0.026 & -- &  22.1 $\pm$ 1.6 & 0.447 $\pm$ 0.031 & 0.195 $\pm$ 0.059 &  0.00 $\pm$ 0.00 \\
   277 &  3.37 &    -- &  -- &  -- &  -- &   4.84 $\pm$ 0.13 &  6.83 $\pm$ 0.15 &  1.412 $\pm$ 0.026 & 0.0753 $\pm$ 0.0047 &  10.9 $\pm$ 0.2 & 1.877 $\pm$ 0.051 & 0.753 $\pm$ 0.017 &  9.99 $\pm$ 0.59 \\
   638 &  6.73 &    7.6 $\pm$  1.3 &  42.9 $\pm$  7.1 &  2.95 $\pm$ 0.49 &  72.1 $\pm$ 12.0 &   2.57 $\pm$ 0.05 & 14.56 $\pm$ 0.17 &  5.673 $\pm$ 0.103 & 0.2445 $\pm$ 0.0044 &  13.8 $\pm$ 0.1 & 3.538 $\pm$ 0.073 & 1.604 $\pm$ 0.019 &  6.56 $\pm$ 0.10 \\
   623 &  5.70 &   13.4 $\pm$  1.6 &  39.1 $\pm$  4.5 &  2.20 $\pm$ 0.25 &  49.9 $\pm$  5.8 &   6.06 $\pm$ 0.14 & 17.72 $\pm$ 0.18 &  2.924 $\pm$ 0.064 & 0.2263 $\pm$ 0.0034 &  19.4 $\pm$ 0.2 & 1.499 $\pm$ 0.035 & 1.951 $\pm$ 0.020 &  8.62 $\pm$ 0.11 \\
    45 &  2.04 &   30.9 $\pm$  3.6 &  56.8 $\pm$  6.6 &  6.40 $\pm$ 0.74 &  56.5 $\pm$  6.6 &   4.83 $\pm$ 0.04 &  8.87 $\pm$ 0.08 &  1.836 $\pm$ 0.010 & 0.0883 $\pm$ 0.0012 &  12.6 $\pm$ 0.1 & 1.879 $\pm$ 0.017 & 0.977 $\pm$ 0.008 & 11.06 $\pm$ 0.13 \\
   214 &  2.25 &   49.8 $\pm$ 10.1 &   9.3 $\pm$  1.9 &  4.84 $\pm$ 0.98 &   7.8 $\pm$  1.6 &  10.30 $\pm$ 0.09 &  1.92 $\pm$ 0.02 &  0.186 $\pm$ 0.002 & 0.0160 $\pm$ 0.0010 &  12.0 $\pm$ 0.1 & 0.881 $\pm$ 0.007 & 0.211 $\pm$ 0.003 & 13.16 $\pm$ 0.80 \\
    33 &  1.32 &   50.3 $\pm$  9.6 &  31.0 $\pm$  5.9 &  7.11 $\pm$ 1.35 &  27.0 $\pm$  5.2 &   7.09 $\pm$ 0.08 &  4.37 $\pm$ 0.04 &  0.617 $\pm$ 0.005 & 0.0380 $\pm$ 0.0011 &  11.0 $\pm$ 0.1 & 1.281 $\pm$ 0.015 & 0.481 $\pm$ 0.005 & 12.66 $\pm$ 0.37 \\
    42 &  1.36 &   53.9 $\pm$ 16.5 &  25.8 $\pm$  7.9 &  5.35 $\pm$ 1.63 &  18.3 $\pm$  5.8 &  10.09 $\pm$ 0.15 &  4.82 $\pm$ 0.09 &  0.478 $\pm$ 0.009 & 0.0342 $\pm$ 0.0029 &  14.4 $\pm$ 0.2 & 0.900 $\pm$ 0.013 & 0.531 $\pm$ 0.010 & 15.51 $\pm$ 1.33 \\
    32 &  1.28 &   44.3 $\pm$  9.3 &  16.2 $\pm$  3.4 &  2.94 $\pm$ 0.62 &   7.2 $\pm$  1.7 &  15.07 $\pm$ 0.44 &  5.51 $\pm$ 0.17 &  0.366 $\pm$ 0.006 & 0.0246 $\pm$ 0.0026 &  20.1 $\pm$ 0.6 & 0.603 $\pm$ 0.018 & 0.607 $\pm$ 0.018 & 24.62 $\pm$ 2.50 \\
   251 &  5.10 &   24.2 $\pm$  4.0 &  33.8 $\pm$  5.5 &  3.94 $\pm$ 0.68 &  31.4 $\pm$  5.2 &   6.15 $\pm$ 0.33 &  8.59 $\pm$ 0.46 &  1.398 $\pm$ 0.015 & 0.0798 $\pm$ 0.0048 &  13.6 $\pm$ 0.7 & 1.477 $\pm$ 0.079 & 0.946 $\pm$ 0.051 & 11.86 $\pm$ 0.34 \\
    62 &  1.72 &   -- &   -- &  -- &   -- &  11.86 $\pm$ 0.11 &  2.06 $\pm$ 0.03 &  0.174 $\pm$ 0.002 & 0.0137 $\pm$ 0.0015 &  13.7 $\pm$ 0.1 & 0.766 $\pm$ 0.007 & 0.227 $\pm$ 0.003 & 16.50 $\pm$ 1.79 \\
    27 &  0.712 &   -- &  -- &  -- &  -- &  10.31 $\pm$ 0.30 &  3.02 $\pm$ 0.17 &  0.293 $\pm$ 0.014 & 0.0167 $\pm$ 0.0036 &  13.1 $\pm$ 0.4 & 0.880 $\pm$ 0.026 & 0.333 $\pm$ 0.018 & 19.95 $\pm$ 4.35 \\
   301 &  1.53 &   70.7 $\pm$ 11.5 &  13.6 $\pm$  2.2 &  6.70 $\pm$ 1.09 &  10.4 $\pm$  1.9 &  10.56 $\pm$ 0.17 &  2.03 $\pm$ 0.04 &  0.192 $\pm$ 0.004 & 0.0155 $\pm$ 0.0014 &  12.4 $\pm$ 0.2 & 0.860 $\pm$ 0.014 & 0.223 $\pm$ 0.005 & 14.39 $\pm$ 1.35 \\
     4 &  1.53 &   43.0 $\pm$  6.1 &  20.6 $\pm$  3.1 &  5.68 $\pm$ 0.80 &  16.3 $\pm$  2.4 &   7.56 $\pm$ 0.06 &  3.63 $\pm$ 0.18 &  0.480 $\pm$ 0.024 & 0.0288 $\pm$ 0.0009 &  10.9 $\pm$ 0.2 & 1.201 $\pm$ 0.010 & 0.400 $\pm$ 0.020 & 13.89 $\pm$ 0.80 \\
    79 &  0.747 &   67.4 $\pm$ 12.8 &  56.5 $\pm$ 10.7 & 10.06 $\pm$ 1.91 &  51.3 $\pm$  9.8 &   6.70 $\pm$ 0.07 &  5.61 $\pm$ 0.04 &  0.837 $\pm$ 0.007 & 0.0510 $\pm$ 0.0009 &  11.7 $\pm$ 0.1 & 1.354 $\pm$ 0.013 & 0.618 $\pm$ 0.004 & 12.12 $\pm$ 0.20 \\
   87E &  1.12 &   67.3 $\pm$ 17.5 &  16.6 $\pm$  4.3 &  6.91 $\pm$ 1.80 &  11.4 $\pm$  3.1 &   9.74 $\pm$ 0.10 &  2.41 $\pm$ 0.04 &  0.247 $\pm$ 0.004 & 0.0165 $\pm$ 0.0014 &  11.9 $\pm$ 0.1 & 0.933 $\pm$ 0.009 & 0.265 $\pm$ 0.004 & 16.08 $\pm$ 1.38 \\
   302 &  2.09 &   55.9 $\pm$ 11.3 &  15.9 $\pm$  3.2 &  5.39 $\pm$ 1.09 &  15.1 $\pm$  3.0 &  10.37 $\pm$ 0.07 &  2.94 $\pm$ 0.03 &  0.284 $\pm$ 0.002 & 0.0279 $\pm$ 0.0007 &  13.0 $\pm$ 0.1 & 0.875 $\pm$ 0.006 & 0.324 $\pm$ 0.003 & 11.61 $\pm$ 0.29 \\
   702 &  3.27 &   -- &   -- &  -- &   -- &  15.15 $\pm$ 1.06 &  1.75 $\pm$ 0.44 &  0.115 $\pm$ 0.029 & 0.0435 $\pm$ 0.0135 &  16.2 $\pm$ 1.2 & 0.599 $\pm$ 0.042 & 0.192 $\pm$ 0.049 &  4.42 $\pm$ 1.73 \\
    95 &  2.34 &   67.8 $\pm$ 15.5 &  29.7 $\pm$  6.8 &  7.88 $\pm$ 1.80 &  22.8 $\pm$  5.3 &   8.61 $\pm$ 0.11 &  3.77 $\pm$ 0.06 &  0.438 $\pm$ 0.006 & 0.0289 $\pm$ 0.0016 &  12.0 $\pm$ 0.1 & 1.055 $\pm$ 0.014 & 0.415 $\pm$ 0.007 & 14.36 $\pm$ 0.82 \\
   710 &  3.10 &   29.5 $\pm$  6.0 &   9.7 $\pm$  2.0 &  3.51 $\pm$ 0.71 &   8.7 $\pm$  1.9 &   8.43 $\pm$ 0.14 &  2.76 $\pm$ 0.05 &  0.327 $\pm$ 0.005 & 0.0249 $\pm$ 0.0019 &  10.9 $\pm$ 0.2 & 1.077 $\pm$ 0.018 & 0.304 $\pm$ 0.005 & 12.20 $\pm$ 0.94 \\
   88W &  2.52 &  -- & -- & -- &  -- &   5.54 $\pm$ 0.06 &  6.42 $\pm$ 0.08 &  1.159 $\pm$ 0.010 & 0.0585 $\pm$ 0.0014 &  11.3 $\pm$ 0.1 & 1.639 $\pm$ 0.019 & 0.707 $\pm$ 0.008 & 12.09 $\pm$ 0.27 \\
   691 &  3.29 &   36.2 $\pm$  8.0 &  68.6 $\pm$ 15.2 &  7.07 $\pm$ 1.57 &  74.5 $\pm$ 16.5 &   5.12 $\pm$ 0.06 &  9.70 $\pm$ 0.09 &  1.895 $\pm$ 0.019 & 0.1053 $\pm$ 0.0016 &  13.4 $\pm$ 0.1 & 1.775 $\pm$ 0.022 & 1.068 $\pm$ 0.010 & 10.14 $\pm$ 0.13 \\
   651 &  4.77 &    -- &  -- &  -- &  -- &   7.31 $\pm$ 0.25 & 11.64 $\pm$ 0.26 &  1.594 $\pm$ 0.047 & 0.1089 $\pm$ 0.0055 &  17.1 $\pm$ 0.4 & 1.243 $\pm$ 0.043 & 1.282 $\pm$ 0.028 & 11.77 $\pm$ 0.55 \\
  740W &  4.12 &   -- &  -- &  -- &   -- &   8.61 $\pm$ 0.18 &  3.50 $\pm$ 0.07 &  0.406 $\pm$ 0.008 & 0.0174 $\pm$ 0.0015 &  11.9 $\pm$ 0.2 & 1.055 $\pm$ 0.022 & 0.385 $\pm$ 0.008 & 22.15 $\pm$ 1.86 \\

\hline
\end{tabular}
\end{minipage}
\end{table}
\end{landscape}

%% file: tableA4.tex

\setcounter{table}{3}
\begin{table}
\centering
\begin{minipage}{145mm}
\caption{Blue compact dwarf galaxy results.}
\vskip-0.1truein
\begin{tabular}{@{}cccc@{}}
\hline 
~ & \multicolumn{3}{c}{$T_e$~\&~$N_e$~from W08, table 2} \\
  Object & Ne$^{++}$/Ne$^+$ & S$^{3+}$/S$^{++}$ &  Ne/S \\
\hline
      Haro11 &  2.03 $\pm$ 0.02 &  0.166 $\pm$ 0.001 &  19.6 $\pm$ 0.1 \\
     NGC1140 &  2.16 $\pm$ 0.04 &  0.096 $\pm$ 0.001 &  16.8 $\pm$ 0.2 \\
     NGC1569 &  7.29 $\pm$ 0.08 &  0.339 $\pm$ 0.002 &  14.3 $\pm$ 0.1 \\
      IIZw40 & 11.73 $\pm$ 0.22 &  0.708 $\pm$ 0.007 &  11.3 $\pm$ 0.1 \\
     UGC4274 &  0.94 $\pm$ 0.01 &  0.052 $\pm$ 0.002 &  13.1 $\pm$ 0.2 \\
       IZw18 &  3.27 $\pm$ 0.39 &  0.413 $\pm$ 0.044 &  14.3 $\pm$ 1.1 \\
     Mrk1450 &  4.66 $\pm$ 0.11 &  0.284 $\pm$ 0.002 &  14.6 $\pm$ 0.1 \\
       UM461 & 11.39 $\pm$ 1.43 &  1.060 $\pm$ 0.040 &  13.8 $\pm$ 0.3 \\
     Mrk1499 &  1.77 $\pm$ 0.03 &  0.116 $\pm$ 0.002 &  16.9 $\pm$ 0.2 \\
\hline
\end{tabular}
\end{minipage}
\end{table}

%% file: tableA5.tex

\setcounter{table}{4}
\begin{table*}
\centering
\begin{minipage}{200mm}
\caption{NGC 3603, 30 Doradus, N66 results.}
\vskip-0.1truein
\begin{tabular}{@{}cccccccccc@{}}
\hline 
  Object & \underbar{Ne$^+$} & \underbar{Ne$^{++}$} & \underbar{Ne} & \underbar{S$^{++}$} & \underbar{S$^{3+}$} &  \underbar{S} &  \underbar{Ne$^{++}$} & \underbar{S$^{3+}$} &  \underbar{Ne} \\
& H$^+$ & H$^+$ & H & H$^+$  & H$^+$ & H & Ne$^+$ & S$^{++}$ & S \\
 & $($$\times${10$^{-6}$}$)$ & $($$\times${10$^{-6}$}$)$ & $($$\times${10$^{-6}$}$)$ & $($$\times${10$^{-6}$}$)$ & $($$\times${10$^{-7}$}$)$ & $($$\times${10$^{-6}$}$)$ &  \\

\hline
NGC 3603\#3 &  54.29 & 48.05 & 102.34 & 5.23 &  3.6 &  5.58 &  0.89 &  0.07 &  18.3 \\
NGC 3603\#4 &  67.11 & 25.64 &  92.76 & 8.68 &  1.9 &  8.88 &  0.38 &  0.02 &  10.4 \\
NGC 3603\#5 &  42.57 & 60.27 & 102.84 & 9.02 &  6.5 &  9.67 &  1.42 &  0.07 &  10.6 \\
\\
 30 Dor\#2 &  19.01 & 48.77 &  67.78 & 5.76 &  4.0 &  6.17 &  2.57 &  0.07 &  11.0 \\
 30 Dor\#3 &  11.48 & 56.36 &  67.84 & 5.31 &  8.1 &  6.12 &  4.91 &  0.15 &  11.1 \\
 30 Dor\#4 &   7.21 & 64.23 &  71.44 & 4.80 & 14.7 &  6.27 &  8.91 &  0.31 &  11.4 \\
 30 Dor\#5 &  23.03 & 53.12 &  76.16 & 6.59 &  5.2 &  7.11 &  2.31 &  0.08 &  10.7 \\
 30 Dor\#6 &  20.32 & 62.42 &  82.74 & 7.05 &  7.0 &  7.75 &  3.07 &  0.10 &  10.7 \\
 30 Dor\#7 &   9.11 & 56.90 &  66.01 & 4.73 & 10.0 &  5.73 &  6.25 &  0.21 &  11.5 \\
 30 Dor\#8 &  28.36 & 33.41 &  61.76 & 5.18 &  2.3 &  5.42 &  1.18 &  0.05 &  11.4 \\
30 Dor\#10 &  40.55 & 21.19 &  61.74 & 5.46 &  2.2 &  5.67 &  0.52 &  0.04 &  10.9 \\
30 Dor\#11 &  17.90 & 61.06 &  78.97 & 6.71 &  6.3 &  7.33 &  3.41 &  0.09 &  10.8 \\
30 Dor\#12 &  13.61 & 66.40 &  80.01 & 5.97 & 10.7 &  7.04 &  4.88 &  0.18 &  11.4 \\
30 Dor\#13 &  13.87 & 51.68 &  65.55 & 5.59 &  6.3 &  6.22 &  3.73 &  0.11 &  10.5 \\
30 Dor\#14 &  12.86 & 62.85 &  75.71 & 5.35 &  9.5 &  6.29 &  4.89 &  0.18 &  12.0 \\
30 Dor\#15 &  12.97 & 73.45 &  86.42 & 6.40 & 10.8 &  7.47 &  5.66 &  0.17 &  11.6 \\
30 Dor\#16 &   7.34 & 61.71 &  69.06 & 4.94 & 12.1 &  6.15 &  8.40 &  0.24 &  11.2 \\
30 Dor\#17 &  34.75 & 20.72 &  55.47 & 4.48 &  1.2 &  4.60 &  0.60 &  0.03 &  12.1 \\
\\
   N66\#1 &   3.97 & 26.98 &  30.95 & 2.65 &  6.1 &  3.27 &  6.79 &  0.23 &   9.5 \\
   N66\#2 &   3.89 & 21.76 &  25.65 & 2.11 &  5.0 &  2.61 &  5.59 &  0.24 &   9.8 \\
   N66\#5 &   3.23 & 17.94 &  21.17 & 1.71 &  3.3 &  2.05 &  5.55 &  0.20 &  10.3 \\
   N66\#6 &   7.96 & 19.86 &  27.82 & 2.25 &  4.6 &  2.71 &  2.50 &  0.20 &  10.3 \\
   N66\#7 &   1.60 & 12.22 &  13.82 & 1.17 &  2.4 &  1.41 &  7.65 &  0.20 &   9.8 \\
   N66\#8 &   6.34 & 21.03 &  27.37 & 2.21 &  3.5 &  2.56 &  3.32 &  0.16 &  10.7 \\
   N66\#9 &   7.05 & 21.07 &  28.11 & 2.24 &  3.0 &  2.54 &  2.99 &  0.13 &  11.1 \\
  N66\#10 &   6.34 & 18.88 &  25.23 & 1.99 &  2.7 &  2.26 &  2.98 &  0.14 &  11.1 \\
  N66\#11 &   7.93 & 23.51 &  31.44 & 2.59 &  3.4 &  2.92 &  2.97 &  0.13 &  10.8 \\
  N66\#12 &   2.31 & 17.93 &  20.23 & 1.35 &  4.9 &  1.84 &  7.77 &  0.36 &  11.0 \\
  N66\#13 &   3.65 & 17.99 &  21.64 & 1.81 &  3.3 &  2.14 &  4.92 &  0.18 &  10.1 \\

\hline
\end{tabular}
\end{minipage}
\end{table*}

%% file: paper.bbl
\begin{thebibliography}{99}

\bibitem[Asplund et al.(2009)]{asplund2009} Asplund M., Grevesse N., Sauval A. J., Scott, P., 2009, ARA\&A, 47, 481


\bibitem[Bouret et al.(2003)]{bouret2003} Bouret J.-C., Lanz T., Hillier D. J., Heap S. R., Hubeny I., Lennon D. J., Smith L. J., Evans C. J., 2003, ApJ, 595, 1182

\bibitem[Bresolin \& Kennicutt(2002)]{bresolin2002} Bresolin F., Kennicutt R. C.\ Jr., 2002, ApJ, 572, 838

\bibitem[Chiappini et al.(2001)]{chiappini2001} Chiappini C., Matteucci F., Romano D., 2001, ApJ, 554, 1044

\bibitem[Chiappini et al.(2003]{chiappini2003} Chiappini C., Romano D., Matteucci F., 2003, MNRAS, 339, 63

\bibitem[Dale et al.(2009)]{dale} Dale D. A., et al., 2009, ApJ, 693, 1821

\bibitem[Del Zanna \& Badnell(2016)]{delzanna16} {Del Zanna G., Badnell N. R., 2016, MNRAS, 456, 3720}

\bibitem[Dufour et al.(1980)]{dufour} Dufour R. J., Talbot R. J.\ Jr., Jensen E. B., Shields G. A., 1980, ApJ, 236, 119

\bibitem[Esteban et al.(2004)]{esteban} Esteban C., Peimbert M., Garc\'ia-Rojas J., Ruiz M.T., Peimbert A., Rodr\'iguez M., 2004, MNRAS, 355, 229

\bibitem[Galametz et al.(2010)]{galametz2010} Galametz M., Madden S. C., Galliano F., Hony S., Sauvage M., Pohlen M., Bendo G. J., Auld R., Baes M., Barlow M.J., 2010 A\&A 518, L5

\bibitem[Garc\'ia-Rojas et al.(2006)]{garcia-rojas} Garc\'ia-Rojas J., Esteban C., Peimbert M., Costado M. T., Rodr\'iguez M., Peimbert A., Ruiz M. T., 2006, MNRAS, 368, 253

\bibitem[Garc\'ia-Rojas et al.(2016)]{garcia-rojas} {Garc\'ia-Rojas J., Pe\~na M., Flores-Dur\'an S., Hern\'andez-Mart\'inez L., 2016, A\&A, 586, 59}

\bibitem[Grieve et al.(2014)]{grieve} Grieve M. F. R., Ramsbottom C. A., Hudson C. E., Keenan F. P., 2014, ApJ, 780, 110

\bibitem[Griffin et al.(2001)]{griffin} Griffin D. C., Mitnik D. M., Badnell N. R., 2001, J.Phys.B, 34, 4401

\bibitem[Helou et al.(1991)]{NED} Helou G., Madore B. F., Schmitz M., Bicay M. D., Wu X., Bennett J., 1991, in Albrecht M. A., Egret D., eds, Databases and On-Line Data in Astronomy. Kluwer, Dordrecht, p. 89

\bibitem[Henry \& Worthey(1999)]{henry99} Henry R. B. C., Worthey G., 1999, PASP, 111, 919

\bibitem[Hern\'andez-Mart\'inez et al.(2009)]{hernandez-martinez} Hern\'andez-Mart\'inez L., Pe\~na M., Carigi L., Garc\'ia-Rojas J., 2009, A\&A, 505, 1027

\bibitem[Higdon et al.(2004)]{higdon} Higdon S. J. U., et al.,  2004, PASP, 116, 975

\bibitem[Hillier \& Miller(1998)]{cmfgen} Hillier D. J., Miller D. L., 1998, ApJ, 496, 407

\bibitem[Hou et al.(2000)]{hou} Hou J. L., Prantzos N., Boissier S., 2000, A\&A, 362, 921

\bibitem[Houck et al.(2004)]{houck2004} Houck J. R., et al., 2004, ApJS, 154, 18

\bibitem[Hubble(1925)]{hubble1925} Hubble E. P., 1925, ApJ, 62, 409

\bibitem[Kennicutt et al.(2003)]{SINGS} Kennicutt R. C. Jr., et al., 2003, PASP, 115, 928

\bibitem[Killen \& Dufour(1982)]{killen} Killen R. M., Dufour R. J., 1982, PASP, 94, 446

\bibitem[Koribalski et al.(2004)]{koribalski} Koribalski B. S., et al., 2004, ApJ, 128, 30

\bibitem[Kubryk et al.(2015)]{kubryk} Kubryk M., Prantzos N., Athanassoula E., 2015, A\&A, 580, A127

\bibitem[Kurucz(1992)]{kurucz} Kurucz R. L., 1992, in Barbuy B., Renzini A., eds, Proc. IAU Symp.\ 149, Stellar Population of Galaxies. Kluwer, Dordrecht, p.\ 225

\bibitem[Lanz \& Hubeny(2003)]{lanz} Lanz T., Hubeny I., 2003, ApJS, 146, 417

\bibitem[Lebouteiller et al.(2008)]{lebouteiller} Lebouteiller V., Bernard-Salas J., Brandl B., Whelan D., Wu Y., Charmandaris V., Devost D., 2008, ApJ, 680, 398 (L08)

\bibitem[Lee et al.(2006)]{lee06} Lee H., Skillman E. D., Venn K. A., 2006, ApJ, 642, 813

\bibitem[Magrini et al.(2007)]{magrini2007} Magrini L., V\'ilchez J. M., Mampaso A., Corradi R. L. M., Leisy P., 2007, A\&A, 470, 865

\bibitem[Martins et al.(2005)]{martins} Martins F., Schaerer D., Hillier D. J., 2005, A\&A, 436, 1049

\bibitem[McLaughlin \& Bell(2000)]{mclaughlin2000} McLaughlin B. M., Bell K. L., 2000, JPhysB, 33, 597

\bibitem[McLaughlin et al.(2011)]{mclaughlin2011} McLaughlin B. M., Lee T.-G., Ludlow J. A., Landi E., Loch S. D., Pindzola M. L., Ballance C. P., 2011, JPhysB, 44, 175206

\bibitem[Melnick et al.(1989)]{melnick} Melnick J., Tapia M., Terlevich R., 1989, A\&A, 213, 89

\bibitem[Mendoza \& Bautista(2014)]{mendoza} Mendoza C., Bautista M.A., 2014, ApJ, 785, 91

\bibitem[Mokiem et al.(2004)]{mokiem} Mokiem M. R., Mart\'{\i}n-Hern\'{a}ndez N.L., Lenorzer A., de Koter A., Tielens A. G. G. M., 2004, A\&A, 419, 319

\bibitem[Patrick et al.(2015)]{patrick} Patrick L. R., Evans C. J., Davies B., Kudritzki R.-P., Gazak J. Z., Bergemann M., Plez B., Ferguson A. M. N., 2015, ApJ, 803, 14 

\bibitem[Pauldrach et al.(2001)]{pauldrach01} Pauldrach A. W. A., Hoffmann T. L., Lennon M., 2001, A\&A, 375, 161

\bibitem[Pauldrach et al.(2012)]{pauldrach12} Pauldrach A. W. A., Vanbeveren D., Hoffmann T. L., 2012, A\&A, 538, A75

\bibitem[Peimbert(2003)]{peimbert2003} Peimbert A., 2003, ApJ, 584, 735

\bibitem[Peimbert et al.(2005)]{peimbert2005} {Peimbert A., Peimbert M., Ruiz M. T. 2005, ApJ, 634, 1056}

\bibitem[Rodr\'{\i}guez \& Rubin(2005)]{rodriguez} Rodr\'{\i}guez M., Rubin R. H., 2005, ApJ, 626, 900

\bibitem[Rolleston et al.(2000)]{rolleston} Rolleston W. R. J., Smartt S. J., Dufton P. L., Ryans R. S. I., 2000, A\&A, 363, 537

\bibitem[Rubin(1985)]{rubin1985} Rubin R. H., 1985, ApJS, 57, 349

\bibitem[Rubin(1989)]{rubin1989} {Rubin R. H., 1989, ApJS, 69, 897}

\bibitem[Rubin et al.(1991)]{rubin1991} Rubin R. H., Simpson J. P., Haas M. R., Erickson E. F., 1991, ApJ, 374, 573

\bibitem[Rubin et al.(1994)]{rubin1994} Rubin R. H., Simpson J. P., Lord S. D., Colgan S. W. J., Erickson E. F., Haas M. R., 1994, ApJ, 420, 772

\bibitem[Rubin et al.(1993)]{rubin1993} Rubin R. H., Dufour R. J., Walter D. K., 1993, ApJ, 413, 242

\bibitem[Rubin et al.(2008)]{rubin2008} Rubin R. H., Simpson J. P., Colgan S. W. J., Dufour R. J., Brunner G., McNabb I. A.,  Pauldrach A. W. A., Erickson E. F., Haas M. R., Citron R. I., 2008, MNRAS, 387, 45 (R08)

\bibitem[Rubin et al.(2007)]{rubin2007} Rubin R. H., Simpson J. P., Colgan S. W. J., Dufour R. J., Ray K. L., Erickson E. F., Haas M. R., Pauldrach A. W. A., Citron R. I., 2007, MNRAS, 377, 1407 (R07)

\bibitem[Rubin et al.(2011)]{rubin2011} Rubin, R. H., Simpson J. P., O'Dell C. R., McNabb I. A., Colgan S. W. J., Zhuge S. Y., Ferland G. J., Hidalgo S. A., 2011, MNRAS, 410, 1320 (R11)

\bibitem[S\'anchez Almeida et al.(2014)]{sanchez-almeida} S\'anchez Almeida J., Elmegreen B. G., Mu\~noz-Tu\~n\'on C., Elmegreen D. M., 2014, A\&ARv, 22, 71 

\bibitem[Saraph \& Storey(1999)]{saraph1999} Saraph H. E., Storey P. J., 1999, A\&AS, 134, 369

\bibitem[Saraph \& Tully(1994)]{saraph94}Saraph H. E., Tully J. A., 1994, A\&AS, 107, 29

\bibitem[Shields(2002)]{shields} Shields G. A., 2002, Rev. Mex. Astron. Astrofis. Ser. Conf., 12, 189

\bibitem[Simpson et al.(2007)]{simp07}{Simpson J. P., Colgan S. W. J., Cotera A. S., Erickson E. F., Hollenbach D. J., Kauffman M. J., Rubin R. H., 2007, ApJ, 670, 1115}

\bibitem[Simpson et al.(2012)]{simpson2012} Simpson J. P., Cotera A. S., Burton M. G., Cunningham M. R., Lo N., Bains I., 2012, MNRAS, 419, 211

\bibitem[Simpson et al.(2004)]{simpson2004} Simpson J. P., Rubin R. H., Colgan S. W. J., Erickson E. F., Haas M. R., 2004, ApJ, 611, 338

\bibitem[Simpson et al.(1998)]{simpson1998} Simpson J. P., Witteborn F. C., Price S. D., Cohen M., 1998, ApJ, 508, 268

\bibitem[Smith et al.(2007)]{smith2007} Smith J. D. T., Armus L., Dale D. A., Roussel H., Sheth K., Buckalew B. A., Jarrett T. H., Helou G., Kennicutt R. C. Jr., 2007, PASP, 119, 1113

\bibitem[Sternberg et al.(2003)]{sternberg} Sternberg A., Hoffmann T. L., Pauldrach A. W. A., 2003, ApJ, 599, 1333

\bibitem[Storey \& Hummer(1995)]{storey} Storey P. J., Hummer D. G., 1995, MNRAS, 272, 41

\bibitem[Tapia et al.(2001)]{tapia} Tapia M., Bohigas J., P\'erez B., Roth M., Ruiz M. T., 2001, Rev. Mex. Astron. Astrofis., 37, 39

\bibitem[Tayal(2000)]{tayal2000} {Tayal S. S., 2000, ApJ, 530, 1091}

\bibitem[Tayal \& Gupta(1999)]{tayal1999} Tayal S. S., Gupta G. P., 1999, ApJ, 526, 544

\bibitem[Tsamis \& Walsh(2011)]{tsamis2011} Tsamis Y. G., Walsh J. R., 2011, MNRAS, 417, 2072

\bibitem[Weber et al.(2015)]{weber15} Weber J. A., Pauldrach A. W. A., Hoffmann T. L., 2015, A\&A, 583, A63

\bibitem[Werner et al.(2004)]{wer04} Werner M. W., et al., 2004, ApJS, 154, 1

\bibitem[Wu et al.(2008)]{wu2008} Wu Y., Bernard-Salas J., Charmandaris V., Lebouteiller V., Hao L., Brandl B. R., Houck J. R., 2008, ApJ, 673, 193 (W08)

\end{thebibliography}
